\documentclass[trackchanges,twocolumn,astrosymb]{aastex701}

\newcommand\SgrA{Sgr~A$^{\star}$}

\newcommand\ergs{$\mathrm{erg\,s^{-1}}$}
\newcommand\um{$\mu$m}
\begin{document}

\title{Infrared Line Diagnostics Fail to Constrain Sgr A*'s UV Output}

\author[orcid=0000-0001-9641-6550,sname='Balakrishnan']{Mayura Balakrishnan}
\altaffiliation{Banting Postdoctoral Fellow}
\affiliation{Department of Physics, McGill University, 3600 Rue University, Montréal, Québec, H3A 2T8, Canada}
\affiliation{Trottier Space Institute at McGill, 3550 Rue University, Montréal, Québec, H3A 2A7, Canada}
\email[show]{mayura.balakrishnan@mcgill.ca}

\author[orcid=0000-0002-9156-2249,sname=von Fellenberg]{Sebastiano D. von Fellenberg}
\altaffiliation{Feodor Lynen Fellow}
\affiliation{Canadian Institute for Theoretical Astrophysics, University of Toronto, 60 St. George Street, Toronto, ON M5S 3H8, Canada}
\affiliation{Max Planck Institute for Radioastronomy, auf dem H{\"u}gel 69, Bonn, Germany }
\email{sfellenberg@cita.utoronto.ca}  

\author[0000-0001-6803-2138]{Daryl Haggard}
\affiliation{Department of Physics, McGill University, 3600 Rue University, Montréal, Québec, H3A 2T8, Canada}
\affiliation{Trottier Space Institute at McGill, 3550 Rue University, Montréal, Québec, H3A 2A7, Canada}
\email{daryl.haggard@mcgill.ca}

\author[orcid=0000-0003-3503-3446,sname=Michail, gname=Joseph]{Joseph M. Michail}
\altaffiliation{NSF Astronomy \& Astrophysics Postdoctoral Fellow}
\affiliation{Center for Astrophysics $|$  Harvard {\&} Smithsonian, 60 Garden St., Cambridge, MA 02138-1516, USA}
\email{joseph.michail@cfa.harvard.edu}

\author[orcid=0000-0001-8921-3624,sname=Ford]{Nicole M. Ford} 
\affiliation{Department of Physics, McGill University, 3600 Rue University, Montréal, Québec, H3A 2T8, Canada}
\affiliation{Trottier Space Institute at McGill, 3550 Rue University, Montréal, Québec, H3A 2A7, Canada}
\email{nicole.ford@mail.mcgill.ca}

\author[0000-0002-5599-4650]{Joseph L. Hora}
\email{jhora@cfa.harvard.edu}
\affiliation{Center for Astrophysics $|$  Harvard {\&} Smithsonian, 60 Garden St., Cambridge, MA 02138-1516, USA}

\author[0000-0002-5635-3345]{Laurent Loinard}
\affiliation{Instituto de Radioastronomía y Astrofísica, Universidad Nacional Autónoma de México, Morelia 58089, México}
\affiliation{Black Hole Initiative at Harvard University, 20 Garden Street, Cambridge, MA 02138, USA}
\email{laurent@astro.unam.mx}

\author[0000-0001-9564-0876]{Sera Markoff}
\affiliation{Anton Pannekoek Institute for Astronomy, University of Amsterdam, Science Park 904, 1098 XH Amsterdam, The Netherlands}
\affiliation{Gravitation and Astroparticle Physics Amsterdam Institute, University of Amsterdam, Science Park 904, 1098 XH 195 196 Amsterdam, The Netherlands}
\affiliation{Institute of Astronomy, University of Cambridge, Madingley Road, Cambridge CB3 0HA, United Kingdom}
\email{S.B.Markoff@uva.nl}

\author[0000-0002-8247-786X]{Joey Neilsen}
\affiliation{Department of Physics, Villanova University, 800 Lancaster Avenue, Villanova, PA 19085, USA}
\email{joseph.neilsen@villanova.edu}

\author[0000-0003-0406-7387]{Giacomo Principe}
\affiliation{Dipartimento di Fisica, Universit\'a di Trieste, I-34127 Trieste, Italy}
\affiliation{Istituto Nazionale di Fisica Nucleare, Sezione di Trieste, I-34127 Trieste, Italy}
\affiliation{INAF Istituto di Radioastronomia, Via P. Gobetti, 101, I-40129 Bologna, Italy}
\email{giacomo.principe@inaf.it}

\author[0000-0001-7134-9005]{Nadeen B. Sabha}
\affiliation{Universit\"{a}t Innsbruck, Institut für Astro- und Teilchenphysik, Technikerstr. 25/8, 6020 Innsbruck, Austria}
\email{Nadeen.Sabha@uibk.ac.at}

\author[]{Howard A. Smith}
\affiliation{Center for Astrophysics $|$  Harvard {\&} Smithsonian, 60 Garden St., Cambridge, MA 02138-1516, USA}
\email{hsmith@cfa.harvard.edu}

\author[0009-0004-8539-3516]{Zach Sumners}
\affiliation{Department of Physics, McGill University, 3600 Rue University, Montréal, Québec, H3A 2T8, Canada}
\affiliation{Trottier Space Institute at McGill, 3550 Rue University, Montréal, Québec, H3A 2A7, Canada}
\email[]{ronald.sumners@mail.mcgill.ca}

% \author[0000-0002-9895-5758]{Steven P. Willner}
% \affiliation{Center for Astrophysics | Harvard \& Smithsonian, 60 Garden Street, Cambridge, MA, USA 02138}
% \email{swillner@cfa.harvard.edu}

\author[0000-0002-2967-790X]{Shuo Zhang}
\affiliation{Department of Physics and Astronomy, Michigan State University, 567 Wilson Rd, East Lansing, MI 48824, USA}
\email{zhan2214@msu.edu}

% \author[0009-0003-9906-2745]{Roychowdhury, Tamojeet}
% \affiliation{Department of Astronomy, University of California Berkeley, Berkeley, CA 94704, USA}
% \email{tamojeet@iitb.ac.in}

% \author[0000-0003-2618-797X]{Witzel, Gunther}
% \affiliation{Max Planck Institute for Radio Astronomy, Bonn \& 53121, Germany}
% \email{gwitzel@mpifr-bonn.mpg.de}

% \author[orcid=0009-0001-1040-4784]{Braden Seefeldt-Gail}
% \affiliation{Canadian Institute for Theoretical Astrophysics, University of Toronto, 60 St. George Street, Toronto, ON M5S 3H8, Canada}
% \email{braden.gail@mail.utoronto.ca}
% \affiliation{Dunlap Institute for Astronomy and Astrophysics, University of Toronto, 50 St. George Street, Toronto, ON M5S 3H4, Canada}

% \author[0000-0001-7801-0362]{Philippov, Alexander }
% \affiliation{University of Maryland, College Park, MD 20742, USA.}
% \email{sashaph@umd.edu}

% \author[0000-0002-7301-3908]{Bart Ripperda}
% \affiliation{Canadian Institute for Theoretical Astrophysics, University of Toronto, 60 St. George Street, Toronto, ON M5S 3H8, Canada.}
% \affiliation{Dunlap Institute for Astronomy and Astrophysics, University of Toronto, 50 St. George Street, Toronto, ON M5S 3H4, Canada.}
% \affiliation{Department of Physics, University of Toronto, 60 St. George Street, Toronto, ON M5S 1A7, Canada.}
% \email{bartripperda@gmail.com}

\email{mayura.balakrishnan@mcgill.ca}
% \author{River Europe}
% \affiliation{University of Heidelberg}
% \email{fakeemail4@google.com}

% \author[0000-0000-0000-0003,sname=Asia,gname=Mountain]{Asia Mountain}
% \altaffiliation{Astrosat Post-Doctoral Fellow}
% \affiliation{Tata Institute of Fundamental Research, Department of Astronomy}
% \email{fakeemail5@google.com}

% \author[0000-0000-0000-0004]{Coral Australia}
% \affiliation{James Cook University, Department of Physics}
% \email{fakeemail6@google.com}

% \author[gname=IceSheet]{Penguin Antarctica}
% \affiliation{Amundsen–Scott South Pole Station}
% \email{fakeemail7@google.com}

\begin{abstract}
\SgrA, the $4 \times 10^{6}\ M_{\odot}$ supermassive black hole at the Galactic Center, exhibits frequent flaring with X-ray luminosities of $L_X \sim 10^{35}$--$10^{36}$ erg s$^{-1}$, while its ultraviolet (UV) emission remains unconstrained due to extreme extinction ($A_V \sim 30$ mag). We use JWST/MIRI time-resolved spectroscopy of the central $0\farcs3$ region to search for mid-infrared emission line variability driven by \SgrA\ flares, comparing the results to \texttt{CLOUDY} photoionization models spanning flare luminosities of $L_{\mathrm{UV}} = 10^{32}$--$10^{39}$ erg s$^{-1}$. We detect no continuum-correlated variability in any mid-infrared line, including [\ion{Fe}{2}] (5.34~$\mu$m), [\ion{Ne}{2}] (12.813~$\mu$m), [\ion{Fe}{2}] (17.936~$\mu$m), and [\ion{S}{3}] (18.713~$\mu$m) over lags of $\sim1$--8~hr. Despite expectations of a flare-driven response, we show that the lack of variability is consistent with the physical conditions in the spatially extended line-emitting gas, where light-crossing timescales of $\sim0.1$--$10$ days and recombination and cooling timescales much longer than the variability timescales suppress any observable response to individual flares. We further find that the modeled infrared spectra are dominated by continuum emission rather than isolated line emission. The brightest predicted lines are intrinsically weak (lower than 10$^{-5}$ mJy), and their detectability is further reduced by the large kinematic broadening expected at the \texttt{CLOUDY}-predicted emitting radii, reducing their contrast against the continuum. Extending the analysis to higher-ionization mid-infrared and near-infrared lines does not improve sensitivity. These results demonstrate that infrared emission lines trace a steady-state radiation field rather than individual flaring events, and therefore infrared line diagnostics cannot be used to constrain the instantaneous UV flux of Sgr A*.
\end{abstract}


\section{Introduction}

At a distance of 8.15 kpc and with a mass of roughly 4.1$\times10^6 M_{\odot}$ \citep{Ghez2008,GRAVITY2019}, the supermassive black hole at the center of the Milky Way, Sagittarius A$^{\star}$ (\SgrA), offers a unique opportunity to investigate the physics of low-luminosity active galactic nuclei (LLAGN) on spatial scales that cannot be resolved in more distant systems. Despite being embedded within a dense nuclear star cluster containing numerous mass-losing stars and streams of molecular gas, \SgrA\ does not have a high accretion rate and is extraordinarily faint. Its bolometric luminosity is $L_{bol} \lesssim10^{36}$ \ergs\ \citep{Yuan2003}, corresponding to an Eddington ratio of $\sim10^{-9}$, many orders of magnitude below that expected from standard radiatively efficient accretion disks \citep{Narayan1998}. The faint end of the Palomar low-luminosity AGN sample reaches nuclear H$\alpha$ luminosities of $L_{\rm H\alpha}\sim10^{37}$ \ergs \citep{Ho2008}, already several orders of magnitude above the quiescent X-ray luminosity of \SgrA. This extreme underluminosity is generally interpreted as evidence that the accretion flow onto \SgrA\ operates in a radiatively inefficient regime. Current constraints favor a broader class of radiatively inefficient accretion flows (RIAFs), in which the gas is hot, optically thin, and geometrically thick, but where significant mass loss occurs through outflows or winds \citep{Narayan1994, Narayan1998, Yuan2003}. Measurements of linear polarization in the submillimeter \citep{Marrone2006} and X-ray constraints on the density profile \citep{Wang2013} limit the accretion rate at small radii to $\dot{M} \lesssim 10^{-7} M_{\odot} \mathrm{yr^{-1}}$, implying that the majority of the inflowing material does not reach the black hole. Constructing a comprehensive broadband spectral energy distribution (SED) for \SgrA\ is therefore essential for understanding the thermodynamic structure of the flow, the efficiency of particle acceleration, and the mechanisms responsible for the observed radiation \citep[see, e.g.,][]{EHT2023}.

While \SgrA\ spends most of its time in a quiescent, low-luminosity state, it exhibits strong variability across multiple wavelengths. X-ray flares reach factors of a few to $\sim10^2$--$10^3$ times the quiescent level, occurring on average $\sim$1 per day \citep[see][and references therein]{Sumners2026,Neilsen2013}, while near-infrared (NIR) emission varies by factors of a few up to $\sim$100, with typical rates of $\sim$4 events per day \citep{Schodel2011,Do2019}. Long-term \textit{Chandra} monitoring has detected dozens of X-ray flares with peak luminosities of $L_X \sim 10^{35}$--$10^{36}$~\ergs\ in the 2--10~keV band and durations ranging from hundreds of seconds to several kiloseconds \citep{Neilsen2013,Ponti2015,Bouffard2019}, with the most extreme events reaching a few $\times 10^{35}$~\ergs\ \citep{Porquet2003,Haggard2019}. Recent systematic analyses further show that flare properties are not self-similar: more luminous flares exhibit harder spectra, increased morphological complexity, and higher fluence for longer durations, implying more efficient energy injection and evolving particle acceleration conditions \citep{Sumners2026}.

Multiwavelength observations show that the X-ray emission traces only a fraction of the total radiative output during flares \citep{Marrone2008}. Spectral modeling of the broadband emission further supports this picture, indicating that the X-ray band captures only a subset of the total energy budget \citep{Markoff2001,Dodds-Eden2009,Boyce2019,Boyce2022}. Applying bolometric corrections of a factor of a few to an order of magnitude therefore implies total flare luminosities of $L_{\mathrm{bol}} \sim 10^{36}$--$10^{37}$~\ergs\ in mm-IR wavelengths. A complete understanding of \SgrA\ flares therefore requires knowledge of the full SED, particularly in the poorly constrained ultraviolet (UV) regime where a significant fraction of the radiative power may emerge in accretion flow models. However, direct observations in the UV are precluded by the extreme interstellar extinction toward the Galactic Center, with $A_V \sim 30$~mag \citep{vonFellenberg2025_extinction,Becklin1968,Becklin1978,Rieke1985, Fritz2011}, corresponding to optical depths that effectively extinguish all UV and optical photons along the line of sight. This creates a substantial observational gap between the NIR and X-ray bands \citep{Melia2001,Drappeau2013}.

Because the UV continuum from \SgrA\ is inaccessible along our line of sight, its luminosity must instead be constrained indirectly through its impact on the surrounding environment. Early studies used the absence of strong dust-heating signatures and far-infrared dust re-radiation to place upper limits on the optical/UV luminosity of $\sim10^{4}$--$10^{5}\ L_{\odot}$ \citep{Falcke1996,Serabyn1997}. Additional scenarios have explored the possibility that past episodes of higher accretion may have left behind remnant (“fossil”) disk structures in the Galactic Center \citep{Nayakshin1998}. Additional constraints on \SgrA's past history come from the lack of strong Compton reflection features or bright X-ray echoes in molecular clouds close to \SgrA, which would be expected if the source sustained a significantly higher UV/X-ray luminosity over extended periods before fading to its current state \citep{Koyama1996, Ponti2010, Clavel2013}. Together, these arguments indicate that despite energetic flaring, the steady-state UV output of \SgrA\ remains modest and must be inferred through secondary tracers rather than direct detection.

A natural time-domain extension of this environmental approach is to search for a delayed response of nearby gas to the variable radiation field of \SgrA. This is conceptually related to reverberation mapping, in which continuum variability from an accreting black hole drives delayed emission-line responses in surrounding gas, allowing light-travel times and gas-response physics to be inferred from correlated variability \citep[see, e.g.,][]{Blandford1982,Bentz2009,Cackett2021}. In classical AGN reverberation mapping, the variable ionizing continuum and broad emission-line response are both observed directly. In the Galactic Center, however, the experiment is adapted to a different regime: the driving UV continuum cannot be observed along our line of sight, and the potentially responding gas is spatially extended, low-density, and embedded in a complex foreground/background environment. A detectable mid-infrared (MIR) line response therefore requires not only sufficient UV flare luminosity, but also light-crossing, recombination, and cooling timescales short enough to preserve variability on the duration of individual flares.

% \textbf{Reverberation mapping provides the standard framework for using time-variable radiation from an unresolved accreting black hole to infer the structure and physical conditions of its surrounding gas. In classical AGN reverberation mapping, variations in the ionizing continuum drive delayed changes in broad emission-line fluxes; the lag between the continuum and line response encodes the characteristic light-travel time to the emitting gas, while the detailed transfer function contains information about the geometry and kinematics of the line-emitting region \citep[see, e.g.][and references therein]{Bentz2009,Blandford1982}. More recent extensions of this technique have applied the same basic principle to continuum-emitting accretion disks, X-ray coronae, and dusty tori, using correlated variability to probe spatial scales that cannot be directly resolved \citep[see][for a recent review]{Cackett2021}. Our experiment is conceptually analogous, but adapted to the Galactic Center: because the UV continuum from \SgrA\ is inaccessible along our line of sight, we search instead for a delayed or induced response in nearby infrared emission lines. In this sense, the mid-infrared gas acts as a potential reverberating screen for otherwise unobservable UV flare emission, although the detectability of such a response depends on whether the relevant light-crossing, recombination, and cooling timescales are short compared to the flare duration.}
Previous searches for line emission from the immediate environment of \SgrA\ have already demonstrated the difficulty of using recombination lines to probe the accretion flow. Motivated in part by reports of broad millimeter hydrogen recombination-line emission from ionized gas within $\sim0\farcs2$--$0\farcs3$ of \SgrA\ \citep{Murchikova2019,YusefZadeh2020}, \citet{Ciurlo2021} searched for a NIR Br$\gamma$ counterpart using 13 years of Keck/OSIRIS data. They found no broad Br$\gamma$ emission associated with the accretion flow and placed a stringent upper limit on the line flux, at least a factor of 80 below the value expected from the reported H30$\alpha$ emission. These results show that even hydrogen recombination lines, which provide a natural tracer of ionized gas, are strongly limited by source confusion, extended foreground/background emission, and the intrinsic faintness of any line component associated with \SgrA.

In this work, we use the unprecedented sensitivity, spatial resolution, and time-resolved spectroscopic capability of the \textit{JWST} Mid-Infrared Instrument (MIRI) to test whether nearby mid-infrared emission lines respond to flaring activity from \SgrA. We first search for intrinsic variability in the observed line light curves and compare the resulting limits with time-dependent photoionization calculations using \texttt{CLOUDY} \citep{Gunasekera2025}. Because this experiment is conceptually related to reverberation mapping, we also perform a complementary lagged cross-correlation and injection$-$recovery analysis of the line and contemporaneous continuum light curves; the methodology and full results are presented in Appendix~\ref{app:crosscorr}. Together, these analyses test whether the surrounding gas can act as a responsive tracer of the otherwise inaccessible UV radiation field of \SgrA.

In Section~\ref{sec:methods}, we describe the construction of emission-line and continuum light curves from the \textit{JWST}/MIRI data and outline the \texttt{CLOUDY} photoionization modeling framework. In Section~\ref{sec:results}, we first show that no robust or physically compelling continuum-correlated variability is detected in the observed \textit{JWST}/MIRI emission-line light curves. We then use time-dependent \texttt{CLOUDY} simulations to show that both the observed and prospective infrared lines are insensitive to hour-timescale flaring from \SgrA. In Section~\ref{sec:discussion}, we interpret these results in the context of gas response timescales and geometric dilution. A complementary search for delayed, continuum-correlated line variability using cross-correlation and injection$-$recovery tests is presented in Appendix~\ref{app:crosscorr}. We summarize our main conclusions in Section~\ref{sec:conclusions}.

\section{Methods} \label{sec:methods}

\subsection{JWST Observations \& Lightcurves} \label{sec:jwst_obs}

\begin{figure}
    \centering
    \includegraphics[width=0.48\textwidth]{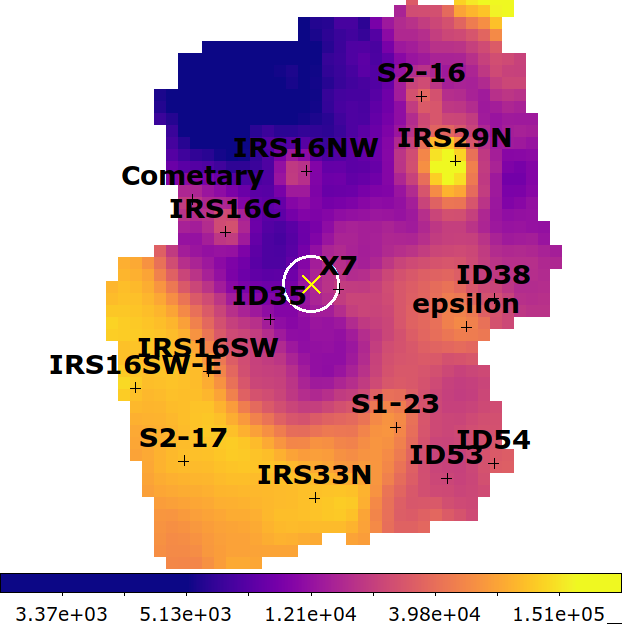}
    \caption{
        \textit{JWST}/MIRI Channel~1 5.6 micron image of the Galactic Center field. Prominent sources are labeled in black. The position of \SgrA\ is marked by the yellow ``X'', and the white circle indicates the 0\farcs3 radius aperture used for spectral extraction. The color scale shows the calibrated surface brightness in units of MJy~sr$^{-1}$.
    }
    \label{fig:jwst_fig}
\end{figure}

\begin{deluxetable}{lclccc}
\tabletypesize{\footnotesize}
\tablecaption{JWST/MIRI MRS time-series observations of \SgrA
\label{tab:obslog}}
\tablehead{
\colhead{Obs.\ ID} & \colhead{Prop.\ ID} & \colhead{UT Date} & \colhead{$N_{\rm int}$} &
\colhead{Span} & \colhead{Cadence} \\
\colhead{} & \colhead{} & \colhead{} & \colhead{(min)} & \colhead{(min)}
}
\startdata
jw04572003001 & 4572 & 2024 Apr 04 & 83  & 117.6 & 1.43 \\
jw04572004001 & 4572 & 2024 Apr 04 & 83  & 117.6 & 1.43 \\
jw04572005001 & 4572 & 2024 Apr 04 & 83  & 117.6 & 1.43 \\
jw04572008001 & 4572 & 2024 Apr 06 & 83  & 117.6 & 1.43 \\
jw04572009001 & 4572 & 2024 Apr 06 & 83  & 117.6 & 1.43 \\
jw04572013001 & 4572 & 2024 Apr 08 & 83  & 117.6 & 1.43 \\
jw04572014001 & 4572 & 2024 Apr 08 & 83  & 117.6 & 1.43 \\
jw04572015001 & 4572 & 2024 Apr 08 & 83  & 117.6 & 1.43 \\
jw04572019001 & 4572 & 2024 Apr 09 & 83  & 117.6 & 1.43 \\
jw04572117001 & 4572 & 2024 Sep 06 & 83  & 117.6 & 1.43 \\
jw04572120001 & 4572 & 2024 Sep 06 & 83  & 117.6 & 1.43 \\
jw07532001001 & 7532 & 2025 Aug 05 & 293 & 418.6 & 1.43 \\
jw07532002001 & 7532 & 2025 Aug 07 & 293 & 418.6 & 1.43 \\
jw07532003001 & 7532 & 2025 Aug 09 & 293 & 418.6 & 1.43 \\
jw07532004001 & 7532 & 2025 Aug 11 & 293 & 418.6 & 1.43 \\
\enddata
\tablecomments{Each row is one MIRI MRS time-series exposure of Sgr~A*, yielding
one light curve per spectral channel. $N_{\rm int}$ is the number of integrations
retained after discarding the first two settling integrations; the span is the
elapsed time from the first to last retained integration; and the cadence is the
mean integration-to-integration sampling. All exposures use the full MRS
wavelength coverage (Channels~1--4, $\sim$4.9--28~$\mu$m). Line and continuum
light curves analyzed in this work are extracted from these cubes as described in
Section~\ref{sec:methods}.}
\end{deluxetable}

\textit{JWST}/MIRI observed Sgr~A* as part of Program PID~4572 \citep[][]{Haggard2023_jwst4572prop, Hora2023_proposal3324} and Program PID 7532 \citep{Hora2025JWST}, using the four SHORT gratings of the MIRI Medium Resolution Spectrograph \citep{Wells2015}. This work uses all available data from these programmes, which provide simultaneous mid-infrared spectral coverage in four channels: Ch~1 (4.90--5.74~$\mu$m), Ch~2 (7.51--8.77~$\mu$m), Ch~3 (11.55--13.47~$\mu$m), and Ch~4 (17.70--20.95~$\mu$m).  All of the data were obtained from the Mikulski Archive for Space Telescopes (MAST) at the Space Telescope Science Institute. The specific observations analyzed can be accessed via \dataset[DOI: 10.17909/580m-xg23]{https://doi.org/10.17909/580m-xg23}. Table \ref{tab:obslog} contains information about the exposures used in the program. An example Channel 1 image of the field is shown in Figure~\ref{fig:jwst_fig}, which illustrates the complex Galactic Center environment and the 0\farcs3 aperture used for our spectral extraction. We use notation for identified objects from \citet{Dinh2024}. The data were processed using the
\href{https://github.com/JoeMichail/MRS_LightCurve_Calibration_Pipeline}
{MRS Light Curve Calibration Pipeline}
\citep{vonFellenberg2025,Michail2026}, which is built on version
2.0.0 of the \texttt{JWST} Calibration Pipeline
\citep{bushouse_2026_19558572} using Calibration Reference Data System
PMAP jwst\_1535. The calibrated products were
then reconstructed into time-resolved spectral cubes. We apply mid-infrared extinction measurements \citep{vonFellenberg2025_extinction} and staring-mode aperture corrections \citep[see Appendix A of][]{Michail2026} to obtain absolute, dereddened photometric light curves.

There are four detected lines sensitive to UV-emission that fall within the wavelength windows allowed by our time series data, the fine structure lines: [\ion{Fe}{2}] (5.34~$\mu$m; Ionization Potential (IP)=7.9 eV), [\ion{Ne}{2}] (12.813~$\mu$m; IP=22 eV), [\ion{Fe}{2}] (17.936~$\mu$m; IP=7.9 eV), and [\ion{S}{3}] (18.713~$\mu$m; IP=23 eV).

No separate blank-sky background field was subtracted. In the central Galactic Center, the MIRI field is dominated by spatially structured astrophysical emission from stars, diffuse gas, dust, and unresolved sources, so there is no source-free local background region representative of the \SgrA\ aperture. We therefore use a differential approach in which we measure line variability from continuum-subtracted light curves defined relative to the local spectral continuum in each integration rather than to an independent blank-sky field.

We extract emission-line and continuum lightcurves from the \textit{JWST}/MIRI MRS spectral cubes for each observing epoch. The first step is to construct a one-dimensional spectrum at each integration time step from a small aperture centered on the WCS-derived position of \SgrA. The calibrated surface-brightness cubes are first converted from MJy sr$^{-1}$ to mJy pixel$^{-1}$ using the FITS pixel solid angle. We then rescale the small-aperture spectrum by the ratio of the large-aperture to small-aperture flux, apply the wavelength-dependent aperture correction, and correct for extinction. Interstellar extinction toward \SgrA\ is corrected at each wavelength using the factor $10^{0.4 A_\lambda}$, where $A_\lambda$ is interpolated from the extinction law of \citet{vonFellenberg2025}, with uncertainties estimated from the dispersion over the $10^5$ posterior samples. The extinction correction is constant in time for a given wavelength and therefore does not affect the detection of relative variability within an individual line light curve; it rescales the fluxes to their intrinsic values and propagates an uncertainty in the absolute flux normalization.

The local continuum is estimated from the wings of the spectral window surrounding each emission line and subtracted using the median flux in these regions. This subtraction is performed independently for each integration and removes the local spectral continuum underlying the line, while preserving temporal changes in the line flux. No temporal detrending is applied to the line spectra before measuring the line fluxes. The continuum-subtracted spectrum is then fit with a Gaussian profile within a narrow interval centered on the line wavelength using bounded nonlinear least-squares optimization, and the integrated line flux is computed analytically from the best-fit Gaussian amplitude and width. The first and last three integrations are removed before constructing the final line light curves and variability statistics to mitigate edge effects.

Uncertainties on the integrated line flux are estimated using four complementary methods. First, we propagate the empirical scatter in the surrounding line-free spectral channels into the integrated line flux. Second, we analytically propagate the Gaussian fit-parameter covariance matrix to the integrated line flux. Third, we draw Monte Carlo realizations of the Gaussian line-profile parameters from the fit covariance matrix and recompute the integrated line flux for each draw. Fourth, we perturb each spectrum using the pipeline-propagated variance array, repeat the continuum subtraction and Gaussian fitting, and use the resulting distribution of integrated fluxes as a pipeline-based Monte Carlo uncertainty. For each observation and line, the representative uncertainty for each method is taken as the median over the retained integrations. In addition, we compute the temporal RMS scatter of each measured line-flux light curve about its median. The first four quantities estimate the per-integration measurement uncertainty, while the temporal RMS characterizes the observed scatter of the time series.

To test for intrinsic line variability, we compare the measured temporal variance of each line light curve with the variance expected from the per-integration measurement uncertainties using the normalized excess variance,
\[ \sigma_{\rm NXS}^2 = \frac{S^2-\langle \sigma_{\rm err}^2\rangle}{\bar{F}^{2}} \]
where $S^2$ is the observed light-curve variance, $\langle \sigma_{\rm err}^2\rangle$ is the mean measurement-error variance, and $\bar{F}$ is the mean line flux. For each line and observation, we adopt the largest finite representative uncertainty among the line-free, analytic fit-covariance, fit-covariance Monte Carlo, and pipeline Monte Carlo estimates as a conservative per-point measurement uncertainty. If the normalized excess variance is consistent with zero, the light curve is consistent with measurement scatter alone, and we report the corresponding upper limit on the fractional variability where applicable. This fractional upper limit is converted to an absolute variable line-flux limit and then to a luminosity limit using $d=8.178$~kpc. We also compute the variability limit separately for each uncertainty method and report the range of method-dependent luminosity limits in Table \ref{tab:line_uncertainties}.

We characterize the empirical photometric stability of each epoch using the continuum lightcurves in line-free wavelength regions. For each MIRI channel, we compute the standard deviation of the baseline-subtracted, extinction-corrected continuum lightcurve in mJy. This quantity is not a direct measurement of the integrated emission-line uncertainty, which is estimated separately using the methods described above. Instead, it provides a continuum-based sensitivity scale for detecting faint time-variable signals in each channel. The measured scatter may include residual instrumental systematics, calibration uncertainties, imperfect removal of spatially structured Galactic Center emission, and intrinsic non-flaring continuum variability within the aperture; we therefore treat it as an observed stability limit rather than as a purely instrumental noise model. This scatter is converted to a monochromatic luminosity via $\nu L_\nu = \nu  \sigma_{\mathrm{corr}} \times 10^{-26} \times 4\pi d^2$, where $\nu = c / \lambda_{\mathrm{ch}}$ and $d = 8.178$~kpc \citep{GRAVITY2019}.

\subsection{CLOUDY Photoionization Models} \label{sec:cloudy}

We model the response of the circumnuclear gas surrounding \SgrA\ using \texttt{CLOUDY} \citep[version c25.00;][]{Gunasekera2025}. The gas is represented as a set of 100 concentric, logarithmically spaced 3D spherical shells spanning radii $r = 10 R_g$ ($2 \times 10^{-6}$ pc; significantly smaller than the \textit{JWST} aperture) to $6 \times 10^{4} R_g$ (0.012~pc; 0\farcs3), where $R_g = GM/c^2 \approx 6.1 \times 10^{11}\ \mathrm{cm}$ for a black hole of mass $4.1 \times 10^{6} M_{\odot}$ \citep{GRAVITY2019, Ghez2008}. The outer radius corresponds to the 0\farcs3 extraction aperture used for the \textit{JWST} light curves. Each shell is treated independently, with constant density and temperature evaluated at the geometric mean radius of each shell.

% The radial structure is prescribed to follow a RIAF profile. The electron temperature scales as $T(r) \propto r^{-1}$, normalized to $T = 10^{10.3}~\mathrm{K}$ at $10 R_g$, while the density follows $n(r) \propto r^{-1}$, normalized to $n = 10^{6.5}~\mathrm{cm^{-3}}$ at $10 R_g$. These choices are motivated by X-ray spectral constraints \citep{Wang2013,Balakrishnan2024a}, GRMHD simulations \citep{Ressler2018}, dynamical measurements of the G2 cloud \citep{Gillessen2019}, and the self-similar RIAF solution \citep{Narayan1994}. We adopt this fixed thermodynamic structure rather than allowing \texttt{CLOUDY} to compute thermal equilibrium, as equilibrium solutions produce temperatures up to $\sim 3$ orders of magnitude lower than expected for the inner RIAF.

The radial structure is prescribed to follow a RIAF profile. The electron temperature scales as $T(r) \propto r^{-1}$, normalized to $T = 10^{10.3}~\mathrm{K}$ at $10 R_g$, while the density follows $n(r) \propto r^{-1}$, normalized to $n = 10^{6.5}~\mathrm{cm^{-3}}$ at $10 R_g$ \citep{EHT2023}. 

The temperature profile is motivated by the nearly virial scaling expected for self-similar radiatively inefficient accretion flows \citep{Narayan1994}. The adopted normalization gives a hot inner flow with $T\sim10^{10.3}$ K at tens of gravitational radii and temperatures of order $10^{6}$--$10^{7}$ K near the outer edge of the modeled aperture. The density normalization is chosen to be consistent, to order of magnitude, with independent constraints on the low-density, outflow-dominated accretion environment around \SgrA. Chandra observations of the quiescent X-ray emission require a tenuous hot plasma with substantial mass loss from the inflow \citep{Wang2013,Balakrishnan2024a}, while the G2 drag measurement probes the density at intermediate radii and gives $n\sim4\times10^{3}~\mathrm{cm^{-3}}$ at $\sim10^{3}$ Schwarzschild radii \citep{Gillessen2019}. Extrapolating our adopted $n\propto r^{-1}$ profile to this radius gives densities within a factor of a few of that constraint. The normalization is also broadly consistent with stellar-wind-fed simulations of the Galactic Center accretion flow, which reproduce the diffuse Chandra thermal emission and reach radii of a few hundred $R_g$ \citep{Ressler2018}. We adopt this fixed thermodynamic structure rather than allowing \texttt{CLOUDY} to compute thermal equilibrium, as equilibrium solutions produce temperatures up to $\sim 3$ orders of magnitude lower than expected for the inner RIAF.

We adopt gas-phase metal abundances of $2 Z_{\odot}$ and exclude dust grains using 
\texttt{abundance ISM no grains}, since dust is not expected to survive in the hot plasma 
considered here, whose minimum temperature is $10^{6.4}$ K 
\citep{Corrales2017,Balakrishnan2024a}. In our fiducial model, iron is further depleted to $0.5 Z_{\odot}$. This choice follows the abundance prescription used in \citet{Balakrishnan2024a} and is also motivated by \textit{JWST}/MIRI-MRS modeling of the central-parsec ionized gas, which favors an $\alpha$-enhanced, Fe-depleted abundance pattern \citep{Vermot2025, Ford2026}, with measurements of $\sim 0.7-2.5 Z_{\odot}$. Our adopted overall metallicity of $2 Z_{\odot}$ therefore lies within the range measured for the Minispiral. The adopted metallicity is also consistent with previous infrared fine-structure-line estimates for the inner Galaxy and Galactic Center \citep{Shields1994,Giveon2002}, as well as stellar $\alpha$-element abundance measurements in the central clusters \citep{Najarro2009}. Because the temperature structure in our models is prescribed rather than solved self-consistently, the adopted abundances primarily rescale the line emissivities approximately linearly. As a result, our main conclusions regarding the absence of detectable line variability are insensitive to plausible abundance revisions at the order-of-magnitude level; they are instead set by the recombination, cooling, and light-crossing timescales of the emitting gas. All results are derived from fully converged \texttt{CLOUDY} simulations evaluated across the full luminosity grid.

The quiescent RIAF's radiation field, the input incident radiation field, is constructed from the multiwavelength SED of \citet{Yuan2003}, normalized to a quiescent bolometric luminosity of $L_{\mathrm{bol}} = 10^{35.5}$~\ergs. Because the bremsstrahlung component of this SED likely originates on larger spatial scales than the synchrotron and inverse-Compton emission relevant to the inner region \citep{Yuan2003,Quataert2002}, we tested models excluding this component and found no significant impact on the results presented in Section~\ref{sec:results}. To model flaring activity, we introduce an additional transient radiation component. Our fiducial model adopts a flare SED motivated by NIR observations of \SgrA\ \citep{GRAVITY2021}. The combined quiescent and flare emission is treated as originating from a central point source and propagated through the surrounding medium.

We emphasize that the incident flare spectrum is not fit directly to the observed infrared continuum. Instead, we adopt a range of physically motivated flare spectral energy distributions, normalized to match typical observed luminosities, and use \texttt{CLOUDY} to compute the resulting line and continuum emission. This approach allows us to test whether any plausible radiation field consistent with the observations can produce detectable infrared line variability.

Time-dependent simulations are performed by injecting a one-hour flare as a top hat function at the beginning of the calculation, followed by the flare undergoing a sharp drop in luminosity (by $\sim 10$ dex), and evolving the system for a total duration of ten hours. The outputs for each shell include time-dependent line emissivities, continuum spectra, temperature structure, and heating rates, enabling us to quantify the contribution of different radial zones to the integrated line response. 

For each shell, we separate the gas response from the direct incident radiation field using the \texttt{CLOUDY} ({\rm total}-{\rm incident}) output. The resulting gas-response spectrum contains the reprocessed continuum and line emission produced by the shell, while excluding the directly incident central-source continuum. When constructing a composite SED for comparison to the observed broadband continuum, we sum the ({\rm total}-{\rm incident}) spectra from all shells and add the incident spectrum once. This composite SED is used only as an instantaneous broadband comparison; it is not interpreted as a simultaneous reverberation signal from gas at all radii. The line-response and variability analysis are based on the shell gas-response spectra, for which light-travel delays and finite recombination/cooling times must be considered separately.

The raw \texttt{CLOUDY} line emissivities should not be interpreted as directly observed, unresolved line profiles. In our shell-by-shell implementation, \texttt{CLOUDY} predicts the radial emissivity distribution of each transition, but the kinematic broadening associated with gas at those radii must be estimated separately. For gas at $r \sim 10^3 R_g$, characteristic of several of the modeled high-ionization lines, the Keplerian velocity scale is $v \sim (GM/r)^{1/2} \sim 10^3~{\rm km~s^{-1}}$. Emission from these radii would therefore be spread over $\Delta\lambda/\lambda \sim v/c$, reducing the peak line contrast relative to the continuum. We therefore treat velocity broadening as a model-inferred consequence of the predicted emitting radii.

We emphasize that the \texttt{CLOUDY} calculations are not intended to model the production of the infrared or X-ray flare emission itself, but rather the response of the surrounding circumnuclear gas to an assumed incident radiation field. The flare SED is treated as an input, motivated by observational constraints, and propagated through the ambient medium to evaluate the resulting line and continuum emission. In particular, the model does not attempt to reproduce the observed (and often non-simultaneous) relationship between infrared and X-ray flares, which likely reflects multiple emission mechanisms or parameter regimes. Instead, our goal is to assess whether any plausible flare radiation field can produce detectable infrared line emission or variability in the surrounding gas.

%To ensure that our variability constraints are not dominated by stellar photoionization, we examine high-ionization-potential lines not detected in the MIRI data, in addition to lower ionization-potential lines in the time-series, which are less susceptible to excitation by the ambient radiation field of Wolf--Rayet stars and therefore provide a cleaner probe of variability driven by \SgrA.

\section{Results} \label{sec:results}

In this section, we quantify the absence of variability in the \textit{JWST}/MIRI emission-line light curves and assess the sensitivity of our observations to flare-driven signals. We first show that no robust continuum-correlated variability is detected (with a full cross-correlation analysis done in Appendix \ref{app:crosscorr}) in any mid-infrared line within the central $0\farcs3$ region, and compare these results to time-dependent \texttt{CLOUDY} simulations to determine whether such variability is expected. We then extend the analysis to higher-ionization MIR and NIR transitions, evaluating whether lines formed at smaller radii provide improved sensitivity to the instantaneous radiation field.

\subsection{Limits on Observed Line Variability} \label{sec:observed_variability}

\begin{figure*}
    \centering
    \includegraphics[width=0.98\textwidth]{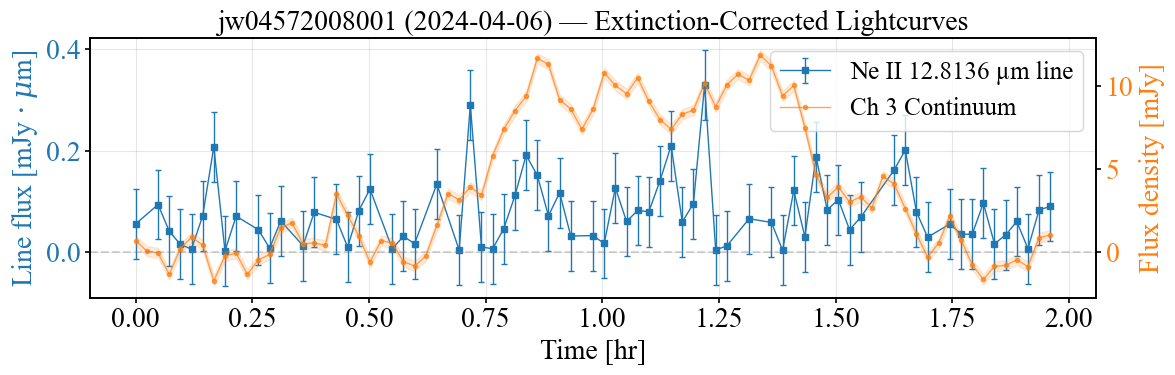}
    \includegraphics[width=0.98\textwidth]{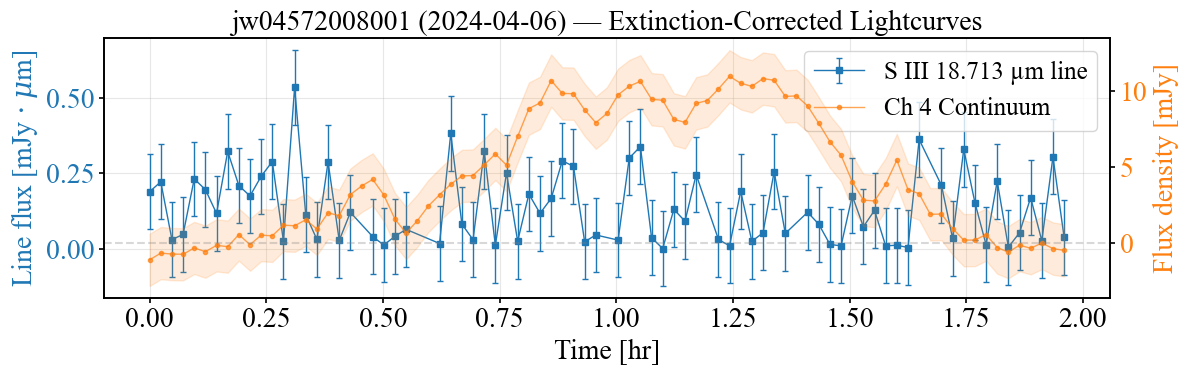}
    \caption{Representative lightcurves for the 2024 observations showing the absence of correlated variability in both [S~III] (18.713 \um) and [Ne~II] (12.813 \um), shown in blue, in response to the \SgrA\ flare described in \citet{vonFellenberg2025} and \citet{Michail2026}. The corresponding continuum lightcurve is plotted in orange. For the emission-line light curves, the error bars show the RMS scatter of the measured line fluxes around their median value and are used as an empirical visual estimate of the line-flux scatter. For the continuum light curves, the shaded bands show the photometric uncertainty propagated from the \textit{JWST} pipeline error array. A full uncertainty analysis and comparison is described in Section \ref{sec:observed_variability}.}
    \label{fig:lightcurves}
\end{figure*}

We construct time-resolved light curves for all mid-infrared emission lines available in the \textit{JWST}/MIRI MRS dataset within the central 0\farcs3 aperture (see Figure \ref{fig:jwst_fig}), including [\ion{Fe}{2}] (5.34~$\mu$m), [\ion{Ne}{2}] (12.813~$\mu$m), [\ion{Fe}{2}] (17.936~$\mu$m), and [\ion{S}{3}] (18.713~$\mu$m). Representative emission-line and continuum lightcurves are shown in Figure~\ref{fig:lightcurves}, corresponding to the flare analyzed in \citet{vonFellenberg2025} and \citet{Michail2026}. Emission-line lightcurves (blue) and continuum lightcurves (orange) show no coherent variability in response to the \SgrA\ flare. We do not detect statistically significant intrinsic variability in the emission-line light curves over the duration of the observations. The line fluctuations also show no visually coherent response to the contemporaneous continuum flare. A quantitative search for delayed continuum-line correlations is presented in Appendix~\ref{app:crosscorr}. While individual integrations occasionally exhibit deviations from the mean flux level, these excursions are not temporally correlated, do not repeat across multiple integrations, and are not observed consistently across different emission lines. This behaviour strongly suggests that they arise from noise fluctuations rather than astrophysical variability. We therefore conclude that the mid-infrared line emission remains consistent with a steady-state signal over the duration of the observations.

We quantify this non-detection using the multi-method line-flux uncertainty analysis described in Section~\ref{sec:methods}. The results are summarized in Table~\ref{tab:line_uncertainties}, which reports the line flux, the four per-integration uncertainty estimates, and the corresponding luminosity upper limit on undetected variability. Specifically, for each line and observing epoch, we compute uncertainties from the empirical scatter in surrounding line-free spectral channels, analytic propagation of the Gaussian fit covariance matrix, Monte Carlo sampling of the Gaussian-fit covariance matrix, and Monte Carlo perturbations using the pipeline-propagated variance array followed by refitting. These four quantities estimate the measurement uncertainty on the integrated line flux, while the temporal RMS characterizes the observed point-to-point scatter of the time series.

For the variability test, we compute the normalized excess variance (see Section \ref{sec:jwst_obs}). We evaluate the variability upper limit separately for each of the four uncertainty prescriptions and report $L_{\rm var,max}$, the largest 3$\sigma$ luminosity upper limit among these methods, as the most conservative limit on undetected line variability. In all cases, the normalized excess variance is consistent with zero within the uncertainties: the observed point-to-point fluctuations are consistent with measurement scatter rather than intrinsic line variability. The resulting limits therefore indicate the maximum line luminosity variation that could remain hidden in the data, rather than a detected variable component.

We note that the representative light curves shown in Figure~\ref{fig:lightcurves} correspond to a 2024 epoch with approximately one hour of post-flare monitoring. The full dataset also includes 2025 epochs with longer temporal coverage, extending to approximately eight hours after the flare. The ten-hour duration of the time-dependent \texttt{CLOUDY} simulations was chosen to encompass this longer observational baseline. Nevertheless, this timescale remains shorter than the light-crossing time of the full 0\farcs3 extraction aperture, whose outer radius corresponds to 0.012 pc, or $\sim$15 light-days. Our observations therefore constrain prompt to several-hour line responses, rather than the full reverberation window of all gas enclosed by the aperture.

As an independent characterization of the continuum stability, we also evaluate the scatter of the continuum light curves in line-free wavelength regions (see Section~\ref{sec:methods}). This continuum-based scatter is not used as the direct uncertainty on the integrated line fluxes, which are estimated separately from the line-profile measurements. Instead, it provides an empirical sensitivity scale for continuum-level variations in each MIRI channel. We find an average uncorrected continuum scatter corresponding to $\sim 3 \times 10^{33}$~\ergs, increasing to $\langle \nu L_\nu \rangle = (7.8 \pm 0.8)\times 10^{33}$~\ergs\ after extinction correction \citep{vonFellenberg2025}, which is higher than most of the estimates derived from errors on the line flux. Modest variations between time segments may reflect a combination of residual instrumental systematics, imperfect subtraction of spatially structured Galactic Center emission, and intrinsic non-flaring continuum variability within the aperture.

\onecolumngrid
\startlongtable
\begin{deluxetable}{lllcccccc}
\tabletypesize{\footnotesize}
\tablecaption{Line fluxes, four-method line-flux uncertainties, and $3\sigma$ variability luminosity upper limits for each epoch and line.}
\tablehead{
  \colhead{Date} & \colhead{Obs ID} & \colhead{Line} & \colhead{$F$} &
  \colhead{$\sigma_{\rm LF}$} & \colhead{$\sigma_{\rm fit}$} & \colhead{$\sigma_{\rm MC}$} &
  \colhead{$\sigma_{\rm pipe}$} & \colhead{$L_{\rm var}^{3\sigma}$} \\
  \colhead{} & \colhead{} & \colhead{} & \colhead{(mJy\,$\mu$m)} &
  \colhead{(mJy\,$\mu$m)} & \colhead{(mJy\,$\mu$m)} & \colhead{(mJy\,$\mu$m)} &
  \colhead{(mJy\,$\mu$m)} & \colhead{(erg\,s$^{-1}$)}
}
\startdata
    2024-04-04 & \texttt{04572003001} & [Fe\,II]\,5.34 & 0.334 & 0.00331 & 0.00157 & 0.00156 & 0.00218 & $8.35\times10^{30}$ \\
    2024-04-04 & \texttt{04572004001} & [Fe\,II]\,5.34 & 0.328 & 0.00332 & 0.00153 & 0.00153 & 0.00211 & $8.39\times10^{30}$ \\
    2024-04-04 & \texttt{04572005001} & [Fe\,II]\,5.34 & 0.335 & 0.00365 & 0.00157 & 0.00156 & 0.00219 & $9.22\times10^{30}$ \\
    2024-04-06 & \texttt{04572008001} & [Fe\,II]\,5.34 & 0.337 & 0.00331 & 0.00156 & 0.00154 & 0.00218 & $8.35\times10^{30}$ \\
    2024-04-06 & \texttt{04572009001} & [Fe\,II]\,5.34 & 0.333 & 0.00355 & 0.00151 & 0.00151 & 0.00208 & $8.95\times10^{30}$ \\
    2024-04-08 & \texttt{04572013001} & [Fe\,II]\,5.34 & 0.334 & 0.00337 & 0.00152 & 0.00153 & 0.00211 & $8.51\times10^{30}$ \\
    2024-04-08 & \texttt{04572014001} & [Fe\,II]\,5.34 & 0.334 & 0.00342 & 0.00155 & 0.00156 & 0.00212 & $8.63\times10^{30}$ \\
    2024-04-08 & \texttt{04572015001} & [Fe\,II]\,5.34 & 0.34 & 0.0034 & 0.00157 & 0.00157 & 0.00213 & $8.57\times10^{30}$ \\
    2024-04-09 & \texttt{04572019001} & [Fe\,II]\,5.34 & 0.34 & 0.00357 & 0.00158 & 0.00159 & 0.00219 & $9.01\times10^{30}$ \\
    2024-09-06 & \texttt{04572117001} & [Fe\,II]\,5.34 & 0.332 & 0.00296 & 0.00148 & 0.00145 & 0.00205 & $7.46\times10^{30}$ \\
    2024-09-06 & \texttt{04572120001} & [Fe\,II]\,5.34 & 0.333 & 0.00315 & 0.00148 & 0.00147 & 0.00206 & $7.94\times10^{30}$ \\
    2025-08-05 & \texttt{07532001001} & [Fe\,II]\,5.34 & 0.334 & 0.0079 & 0.00141 & 0.00141 & 0.00239 & $1.99\times10^{31}$ \\
    2025-08-07 & \texttt{07532002001} & [Fe\,II]\,5.34 & 0.333 & 0.00723 & 0.0014 & 0.0014 & 0.00229 & $1.82\times10^{31}$ \\
    2025-08-09 & \texttt{07532003001} & [Fe\,II]\,5.34 & 0.334 & 0.00737 & 0.00141 & 0.00141 & 0.0023 & $1.86\times10^{31}$ \\
    2025-08-11 & \texttt{07532004001} & [Fe\,II]\,5.34 & 0.336 & 0.0072 & 0.00141 & 0.00142 & 0.0023 & $1.82\times10^{31}$ \\
    2024-04-04 & \texttt{04572003001} & [Ne\,II]\,12.81 & 20.7 & 2.48 & 0.0593 & 0.0587 & 0.0682 & $1.09\times10^{33}$ \\
    2024-04-04 & \texttt{04572004001} & [Ne\,II]\,12.81 & 19.9 & 2.31 & 0.0562 & 0.0563 & 0.0651 & $1.01\times10^{33}$ \\
    2024-04-04 & \texttt{04572005001} & [Ne\,II]\,12.81 & 20.4 & 2.46 & 0.0579 & 0.0579 & 0.0661 & $1.08\times10^{33}$ \\
    2024-04-06 & \texttt{04572008001} & [Ne\,II]\,12.81 & 20.8 & 2.5 & 0.0595 & 0.0593 & 0.0688 & $1.10\times10^{33}$ \\
    2024-04-06 & \texttt{04572009001} & [Ne\,II]\,12.81 & 19.9 & 2.35 & 0.0561 & 0.0561 & 0.0653 & $1.03\times10^{33}$ \\
    2024-04-08 & \texttt{04572013001} & [Ne\,II]\,12.81 & 19.8 & 2.32 & 0.0553 & 0.0553 & 0.063 & $1.02\times10^{33}$ \\
    2024-04-08 & \texttt{04572014001} & [Ne\,II]\,12.81 & 19.7 & 2.31 & 0.0551 & 0.0547 & 0.0622 & $1.01\times10^{33}$ \\
    2024-04-08 & \texttt{04572015001} & [Ne\,II]\,12.81 & 20.7 & 2.49 & 0.0588 & 0.0593 & 0.0683 & $1.09\times10^{33}$ \\
    2024-04-09 & \texttt{04572019001} & [Ne\,II]\,12.81 & 20.4 & 2.47 & 0.0574 & 0.0569 & 0.0662 & $1.08\times10^{33}$ \\
    2024-09-06 & \texttt{04572117001} & [Ne\,II]\,12.81 & 21.4 & 2.49 & 0.0612 & 0.061 & 0.0679 & $1.09\times10^{33}$ \\
    2024-09-06 & \texttt{04572120001} & [Ne\,II]\,12.81 & 21.4 & 2.5 & 0.0612 & 0.061 & 0.0678 & $1.10\times10^{33}$ \\
    2025-08-05 & \texttt{07532001001} & [Ne\,II]\,12.81 & 20.1 & 2.43 & 0.0551 & 0.0552 & 0.0629 & $1.07\times10^{33}$ \\
    2025-08-07 & \texttt{07532002001} & [Ne\,II]\,12.81 & 20.1 & 2.45 & 0.0554 & 0.0555 & 0.0636 & $1.07\times10^{33}$ \\
    2025-08-09 & \texttt{07532003001} & [Ne\,II]\,12.81 & 20.1 & 2.46 & 0.0555 & 0.0554 & 0.0641 & $1.08\times10^{33}$ \\
    2025-08-11 & \texttt{07532004001} & [Ne\,II]\,12.81 & 20.2 & 2.47 & 0.0556 & 0.0551 & 0.0639 & $1.08\times10^{33}$ \\
    2024-04-04 & \texttt{04572003001} & [Fe\,II]\,17.94 & 0.807 & 0.0875 & 0.125 & 0.127 & 0.18 & $4.03\times10^{31}$ \\
    2024-04-04 & \texttt{04572004001} & [Fe\,II]\,17.94 & 0.816 & 0.084 & 0.12 & 0.123 & 0.179 & $3.99\times10^{31}$ \\
    2024-04-04 & \texttt{04572005001} & [Fe\,II]\,17.94 & 0.827 & 0.0864 & 0.127 & 0.128 & 0.193 & $4.32\times10^{31}$ \\
    2024-04-06 & \texttt{04572008001} & [Fe\,II]\,17.94 & 0.783 & 0.0761 & 0.126 & 0.129 & 0.182 & $4.07\times10^{31}$ \\
    2024-04-06 & \texttt{04572009001} & [Fe\,II]\,17.94 & 0.786 & 0.0862 & 0.121 & 0.123 & 0.175 & $3.91\times10^{31}$ \\
    2024-04-08 & \texttt{04572013001} & [Fe\,II]\,17.94 & 0.782 & 0.0827 & 0.118 & 0.122 & 0.169 & $3.78\times10^{31}$ \\
    2024-04-08 & \texttt{04572014001} & [Fe\,II]\,17.94 & 0.812 & 0.0711 & 0.122 & 0.124 & 0.177 & $3.96\times10^{31}$ \\
    2024-04-08 & \texttt{04572015001} & [Fe\,II]\,17.94 & 0.799 & 0.0755 & 0.127 & 0.128 & 0.178 & $3.97\times10^{31}$ \\
    2024-04-09 & \texttt{04572019001} & [Fe\,II]\,17.94 & 0.838 & 0.0633 & 0.127 & 0.128 & 0.185 & $4.14\times10^{31}$ \\
    2024-09-06 & \texttt{04572117001} & [Fe\,II]\,17.94 & 0.683 & 0.0981 & 0.118 & 0.12 & 0.174 & $3.88\times10^{31}$ \\
    2024-09-06 & \texttt{04572120001} & [Fe\,II]\,17.94 & 0.695 & 0.0881 & 0.119 & 0.122 & 0.171 & $3.84\times10^{31}$ \\
    2025-08-05 & \texttt{07532001001} & [Fe\,II]\,17.94 & 0.78 & 0.102 & 0.135 & 0.138 & 0.187 & $4.18\times10^{31}$ \\
    2025-08-07 & \texttt{07532002001} & [Fe\,II]\,17.94 & 0.78 & 0.0961 & 0.131 & 0.134 & 0.187 & $4.18\times10^{31}$ \\
    2025-08-09 & \texttt{07532003001} & [Fe\,II]\,17.94 & 0.756 & 0.115 & 0.129 & 0.131 & 0.184 & $4.12\times10^{31}$ \\
    2025-08-11 & \texttt{07532004001} & [Fe\,II]\,17.94 & 0.762 & 0.0932 & 0.13 & 0.132 & 0.182 & $4.08\times10^{31}$ \\
    2024-04-04 & \texttt{04572003001} & [S\,III]\,18.71 & 16.4 & 0.789 & 0.261 & 0.266 & 0.353 & $1.62\times10^{32}$ \\
    2024-04-04 & \texttt{04572004001} & [S\,III]\,18.71 & 15.9 & 0.643 & 0.26 & 0.259 & 0.362 & $1.32\times10^{32}$ \\
    2024-04-04 & \texttt{04572005001} & [S\,III]\,18.71 & 16.4 & 0.761 & 0.269 & 0.271 & 0.373 & $1.56\times10^{32}$ \\
    2024-04-06 & \texttt{04572008001} & [S\,III]\,18.71 & 15.3 & 0.228 & 0.395 & 0.529 & 0.681 & $1.40\times10^{32}$ \\
    2024-04-06 & \texttt{04572009001} & [S\,III]\,18.71 & 14.7 & 0.18 & 1.38 & 1.96 & 0.638 & $4.02\times10^{32}$ \\
    2024-04-08 & \texttt{04572013001} & [S\,III]\,18.71 & 14.8 & 0.102 & 1.13 & 1.6 & 0.685 & $3.28\times10^{32}$ \\
    2024-04-08 & \texttt{04572014001} & [S\,III]\,18.71 & 14.3 & 0.155 & 0.598 & 0.844 & 0.664 & $1.73\times10^{32}$ \\
    2024-04-08 & \texttt{04572015001} & [S\,III]\,18.71 & 15.2 & 0.584 & 0.252 & 0.261 & 0.438 & $1.20\times10^{32}$ \\
    2024-04-09 & \texttt{04572019001} & [S\,III]\,18.71 & 16.3 & 0.641 & 0.272 & 0.273 & 0.381 & $1.32\times10^{32}$ \\
    2024-09-06 & \texttt{04572117001} & [S\,III]\,18.71 & 14.5 & 0.669 & 0.238 & 0.243 & 0.366 & $1.37\times10^{32}$ \\
    2024-09-06 & \texttt{04572120001} & [S\,III]\,18.71 & 14.8 & 0.697 & 0.243 & 0.247 & 0.379 & $1.43\times10^{32}$ \\
    2025-08-05 & \texttt{07532001001} & [S\,III]\,18.71 & 15.9 & 0.726 & 0.271 & 0.315 & 0.353 & $1.49\times10^{32}$ \\
    2025-08-07 & \texttt{07532002001} & [S\,III]\,18.71 & 15.9 & 1.37 & 0.267 & 0.266 & 0.418 & $2.82\times10^{32}$ \\
    2025-08-09 & \texttt{07532003001} & [S\,III]\,18.71 & 15.9 & 1.42 & 0.268 & 0.268 & 0.419 & $2.92\times10^{32}$ \\
    2025-08-11 & \texttt{07532004001} & [S\,III]\,18.71 & 15.8 & 1.4 & 0.27 & 0.271 & 0.423 & $2.88\times10^{32}$ \\
\enddata
\label{tab:line_uncertainties}
\tablecomments{Fluxes and uncertainties are integrated line quantities in
mJy\,$\mu$m; $L_{\rm var}^{3\sigma}$ is in erg\,s$^{-1}$. $F$ is the
continuum-subtracted, aperture- and extinction-corrected integrated line flux
from the bounded-Gaussian fit. The four uncertainty estimators are:
$\sigma_{\rm LF}$, the empirical scatter in the line-free channels adjacent to the
line; $\sigma_{\rm fit}$, analytic propagation of the Gaussian fit-covariance
matrix; $\sigma_{\rm MC}$, Monte~Carlo resampling of the Gaussian parameters from
the fit covariance; and $\sigma_{\rm pipe}$, Monte~Carlo perturbation of the
spectrum by the propagated pipeline errors followed by refitting. Each column
gives the median over integrations. The $3\sigma$ variability luminosity upper
limit is $L_{\rm var}^{3\sigma}=3\,\sigma_{\rm adopted}$, where $\sigma_{\rm adopted}$
is the largest of the four estimators (converted to a luminosity at the Galactic
Center distance), adopted as a conservative per-integration measurement
uncertainty.}
\end{deluxetable}
\twocolumngrid

\subsection{Time-Dependent CLOUDY Simulations}

We perform fully time-dependent \texttt{CLOUDY} simulations in which a one-hour flare is injected and the subsequent evolution of the gas emission is tracked. Across the full grid of flare luminosities (ranging from 10$^{32}$--10$^{39}$ \ergs), the predicted line emission for MIR fine structure lines remains extremely weak. Predicted emission from observed lines [\ion{Fe}{2}], [\ion{Ne}{2}], and [\ion{S}{3}] remains below the numerical sensitivity threshold across all 100 shells even for the highest luminosity model ($L_{\mathrm{flare}} = 10^{39}$ erg s$^{-1}$). Importantly, the \texttt{CLOUDY} models reproduce the observed continuum normalization in the mid-infrared, indicating that the adopted incident radiation fields are broadly consistent with the measured flux levels. 
%In the time-dependent emissivity output, the only selected MIRI low ionization potential line with nonzero emission is H~I~$7.46~\mu$m, with a characteristic flux of $\sim10^{-4}$ mJy~$\mu$m, while e

\begin{figure*}
    \centering
    \includegraphics[width=0.985\textwidth]{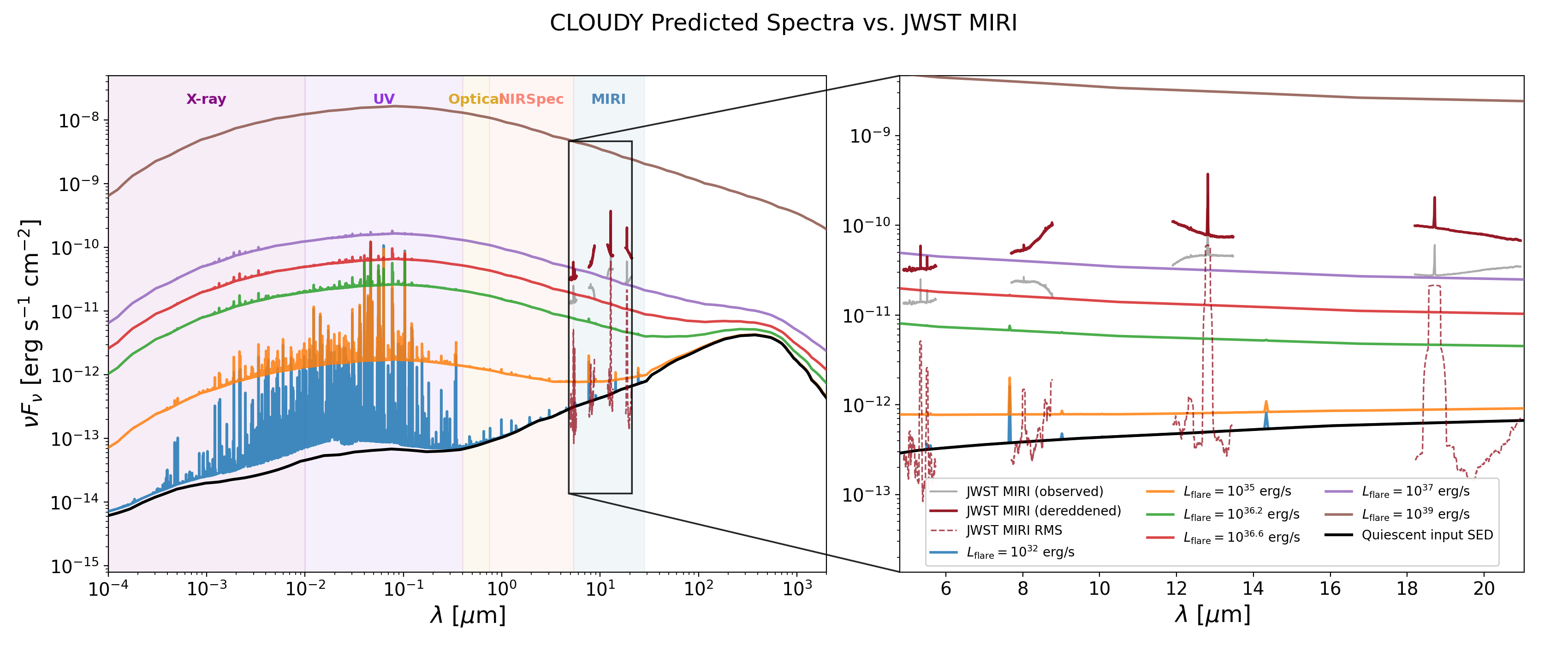}
    \caption{LEFT: Comparison of \texttt{CLOUDY} models to the observed \textit{JWST}/MIRI spectrum of the central $0\farcs3$ around \SgrA. The crimson and gray curves show the MIRI MRS spectra, dereddened and observed, respectively, during the flare window, and the dashed crimson curve shows the local continuum-subtracted RMS. Colored curves show the predicted flare-peak spectra for flare luminosities $L_{\mathrm{flare}} = 10^{32}$--$10^{39}$~erg~s$^{-1}$. The black curve shows the quiescent input SED \citep{Yuan2003}. For most flare models, the emission is continuum dominated, precluding the use of mid-infrared line emission to constrain the UV flux of \SgrA. RIGHT: Zoom-in to MIRI MRS wavelengths}
    \label{fig:cloudy_vs_jwst}
\end{figure*}

Figure~\ref{fig:cloudy_vs_jwst} compares the predicted broadband spectral energy distributions at the flare peak from our \texttt{CLOUDY} models to the observed \textit{JWST}/MIRI MRS spectrum of the central $0\farcs3$ region around \SgrA. The MIRI spectrum (crimson and gray) and its associated continuum-subtracted RMS (dashed crimson) span the $\sim5$–$20~\mu$m range, while the model SEDs extend from the infrared to X-ray energies. The black curve shows the adopted quiescent input SED \citep{Yuan2003}, and the colored curves correspond to models with increasing flare luminosity ($\log L_{\mathrm{flare}} = 32$--39 erg s$^{-1}$).

Across the MIRI wavelength range, the modeled spectra are dominated by smooth continuum emission rather than by discrete line features. Even in the highest-luminosity model, with $\log L_{\rm flare}=39$, the predicted mid-infrared spectrum remains relatively smooth and does not produce strong, isolated emission lines. Increasing the flare luminosity primarily raises the overall normalization of the SED, while leaving the mid-infrared spectral shape broadly similar. In this context, “continuum dominated” refers to the low contrast of the predicted emission lines relative to the underlying continuum. The composite SED includes the direct/transmitted central-source spectrum, shown only for comparison, as well as the gas-reprocessed continuum from the surrounding hot plasma. The gas-reprocessed component is contained in the $({\rm total}-{\rm incident})$ shell output and is dominated by free-free emission under the hot, dust-free conditions adopted here, with other diffuse continuum processes included self-consistently by \texttt{CLOUDY}. We therefore do not interpret the composite SED as a simultaneous observation of the flare continuum and a delayed line response. Instead, it illustrates that, even before accounting for light-travel-time smearing or finite gas-response timescales, the predicted infrared line contrast is intrinsically low.

In comparison to the data, the models lie below the observed MIRI spectrum at low flare luminosities and approach or exceed it only at the highest luminosities. The dashed crimson curves denote the continuum-subtracted RMS of the MIRI data and therefore the effective sensitivity of the observations. No features in either the observed spectrum or its RMS correspond to any structure in the model spectra. This absence of detectable line emission, combined with the continuum-dominated nature of the models, indicates that for plausible flare luminosities the circumnuclear gas does not produce observable MIR line emission nor variability.

\subsection{Extension to Unobserved Lines}
\label{sec:mir_nir}

The wavelength coverage of our time-series observations is limited to the \textit{JWST}/MIRI MRS short channels (sub-band A), which provides spectral coverage in four channels spanning 4.90--5.74~$\mu$m (Ch1), 7.51--8.77~$\mu$m (Ch2), 11.55--13.47~$\mu$m (Ch3), and 17.70--20.95~$\mu$m (Ch4), and primarily probes low- and intermediate-ionization transitions. To evaluate whether alternative diagnostics could provide improved sensitivity to flare-driven variability from \SgrA, we extend our \texttt{CLOUDY} analysis to higher-IP lines in both the MIR and NIR. These coronal transitions are expected to arise at smaller radii, where the gas is more directly exposed to the accretion-powered ionizing continuum and less impacted by the ambient ultraviolet radiation field produced by the surrounding population of massive stars \citep{Ford2026}, including Wolf--Rayet stars in the central parsec \citep{Paumard2006,Martins2007,Lu2013}.

To assess whether line variability should be detectable, it is useful to compare the characteristic flare duration to the relevant gas response timescales. A necessary condition for observable variability is that both the recombination timescale ($t_{\mathrm{rec}}$) and cooling timescale ($t_{\mathrm{cool}}$) be comparable to or shorter than the flare duration ($t_{\mathrm{flare}} \sim 1$ hr), such that the ionization state of the gas can adjust in response to changes in the incident radiation field. In addition, the emitting region must lie within the light-crossing radius ($r \lesssim c\,t_{\mathrm{flare}}$), so that different portions of the gas respond coherently rather than being temporally smeared. If either $t_{\mathrm{rec}} \gg t_{\mathrm{flare}}$ or $t_{\mathrm{cool}} \gg t_{\mathrm{flare}}$, the gas cannot respond on the timescale of the flare and instead reflects a time-averaged radiation field. Similarly, if the bulk of the emission arises from radii $r \gg c\,t_{\mathrm{flare}}$, light-travel-time effects further suppress coherent variability. As we show below, both conditions are violated for all MIR and NIR transitions considered here.

A separate requirement for detectability is that the absolute line flux exceeds the instrumental sensitivity. This distinction is central to the interpretation of Figure~\ref{fig:mir_nir_lc}. Some modeled transitions vary by factors of several to more than an order of magnitude in a relative sense, but their absolute flux densities remain extremely small. We therefore show the model light curves in absolute flux-density units, rather than normalized units, so that the predictions can be compared directly with the sensitivity of \textit{JWST} and with the requirements for future infrared missions. Representative detectable flux-density levels are of order $0.1$--$1$~mJy for MIRI/MRS, depending on wavelength, channel, integration time, source morphology, and background, and of order $10^{-3}$--$10^{-2}$~mJy for favorable NIRSpec/NIRCam observations. These values are order-of-magnitude benchmarks rather than exact limits for the Galactic Center field, but they show that the predicted lines remain far below detectability even when they vary substantially in a fractional sense.

We model these lines using the same time-dependent \texttt{CLOUDY} framework described in Section~\ref{sec:cloudy}, tracking both the radius at which we see peak emission and the line's temporal response to an injected flare. For each transition, we determine the radius of peak emission, local plasma conditions, and characteristic recombination ($t_{\mathrm{rec}}$) and cooling timescales ($t_{\mathrm{cool}}$). The results are summarized in Table~\ref{tab:nir_timescales}, which provides a unified comparison across a broad range of ionization potentials ($\sim50$--$300$~eV) and wavelengths.

Table~\ref{tab:nir_timescales} highlights two key trends. First, higher-IP lines systematically originate at smaller radii ($r \sim 10^{2}$--$10^{3}~R_g$) compared to the MIR low-IP lines ($r \sim 10^{3}$--$10^{4}~R_g$), where densities and temperatures are elevated relative to the outer regions traced by the MIRI lines. Second, despite these more extreme conditions, the recombination and cooling timescales remain long, typically spanning days to years depending on the transition, much longer than the characteristic variability observed in \SgrA\ of $\sim$1 hour.

\subsubsection{High-Ionization MIR Lines}

High-IP MIR lines (e.g., [\ion{Ne}{6}], [\ion{Mg}{7}], [\ion{Si}{7}]) require ionization energies $\gtrsim50$~eV and are therefore not efficiently produced by stellar radiation \citep{Abel2008}. In our models, these transitions arise at $r \sim 10^{2}$--$10^{3}~R_g$, where electron densities reach $n_e \sim 10^{3}$--$10^{4}~\mathrm{cm^{-3}}$. While these conditions reduce recombination timescales relative to the MIR-emitting gas, the response remains slow compared to the flare timescale.

The predicted light curves are shown in Figure~\ref{fig:mir_nir_lc}. These models do not imply that the intrinsic line flux is constant. Indeed, some transitions exhibit substantial relative changes following the injected flare. However, the absolute predicted flux densities remain far below \textit{JWST} sensitivity limits. The high-ionization MIR and NIR light curves in Figure~\ref{fig:mir_nir_lc} are plotted as Gaussian-equivalent peak flux densities assuming a FWHM of $1000~{\rm km~s^{-1}}$. Even for the most extreme flare luminosity considered ($\log L_{\mathrm{flare}} = 39$), the modeled MIR line flux densities remain $\lesssim10^{-8}$--$10^{-5}$~mJy, depending on the transition and on the flux definition. This is several to many orders of magnitude below representative MIRI/MRS detectability levels of $\sim0.1$--$1$~mJy. Thus, even an order-of-magnitude relative variation in these lines would remain observationally inaccessible with \textit{JWST}.

In addition, the dependence on flare luminosity is non-monotonic. As $L_{\mathrm{flare}}$ increases, line emissivities initially rise but subsequently decline due to overionization, which depletes the relevant ionic species (see Figure~\ref{fig:cloudy_vs_jwst}). This saturation further limits the diagnostic utility of these transitions.

\subsubsection{Near-Infrared Coronal Lines}

NIR coronal lines (e.g., [\ion{Si}{9}], [\ion{Si}{7}], [\ion{Mg}{8}]) probe even higher ionization potentials ($\sim50$--$300$~eV) and originate at comparable or smaller radii than the MIR high-IP lines (Table~\ref{tab:nir_timescales}). At these locations, the gas reaches higher densities and temperatures, which in principle should enhance responsiveness to changes in the ionizing field.

However, the results presented in Table~\ref{tab:nir_timescales} demonstrate that even in these inner regions, recombination times remain on the order of days to weeks. As a result, the ionization state cannot adjust on flare timescales, and any response to a short flare is temporally smeared. The model light curves in Figure~\ref{fig:mir_nir_lc} therefore show relative evolution, but not at an absolute level that would be detectable with \textit{JWST}.

The intrinsic fluxes are likewise suppressed. The brightest NIR lines reach peak flux densities of $\sim10^{-7}$~mJy at $\log L_{\mathrm{flare}} = 39$, still orders of magnitude below detectability. For comparison, representative NIRSpec/NIRCam sensitivities are of order $10^{-3}$--$10^{-2}$~mJy for favorable point-source observations, so the modeled NIR coronal lines remain at least $\sim10^{4}$--$10^{5}$ times too faint even in the most optimistic flare models. As in the MIR case, overionization at high luminosities introduces a non-monotonic response, preventing large increases in line strength.

\subsubsection{Implications for Variability Diagnostics}

Table~\ref{tab:nir_timescales} demonstrates that increasing the ionization potential and probing smaller radii does not recover sensitivity to short-timescale variability. Although higher-IP lines trace denser, more strongly irradiated gas, their thermodynamic response remains fundamentally limited by recombination and cooling timescales that are long compared to the duration of \SgrA\ flares. Even the most extreme cases (e.g., [\ion{Si}{9}]) exhibit recombination times of order days, still far exceeding the $\sim$hour duration of typical \SgrA\ flares.

\begin{figure*}
    \centering
    \includegraphics[width=0.98\textwidth]{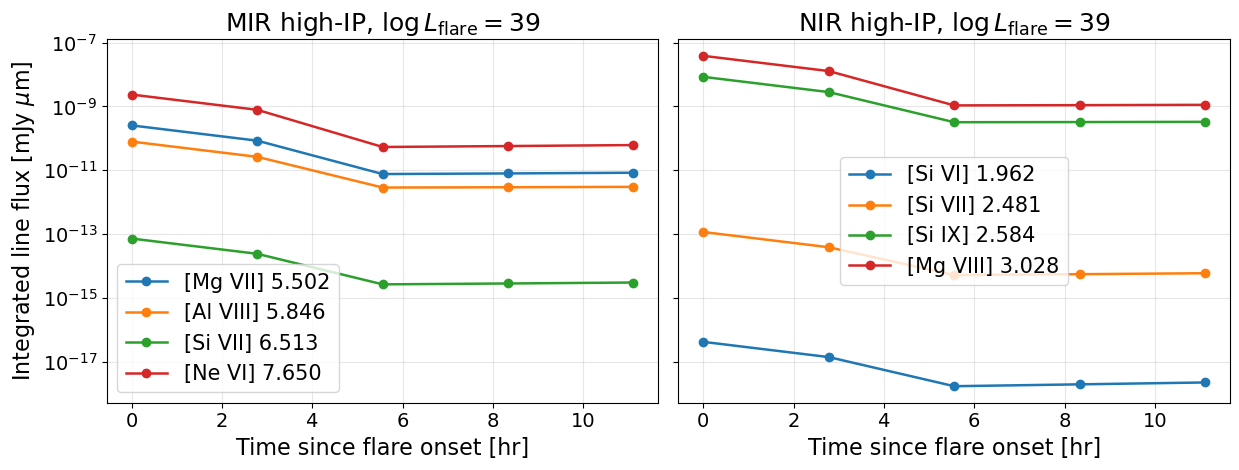}
    \caption{Predicted time-dependent line fluxes for representative high-ionization MIR (left) and NIR (right) transitions at $\log L_{\mathrm{flare}} = 39$, computed with \texttt{CLOUDY}. The flare is injected for the first hour of the simulation. Fluxes are shown as equivalent Gaussian peak flux densities assuming a FWHM of $1000~\mathrm{km~s^{-1}}$. The curves are shown in absolute flux-density units to allow direct comparison with observational sensitivity limits and with the requirements for future infrared missions. Some transitions vary by factors of several to more than an order of magnitude in a relative sense; however, their absolute flux densities remain extremely faint ($\lesssim 10^{-5}$ mJy). For comparison, representative \textit{JWST} sensitivity levels are $\sim0.1$--$1$ mJy for MIRI/MRS and $\sim10^{-3}$--$10^{-2}$ mJy for favorable NIRSpec/NIRCam observations, depending on wavelength, observing setup, source morphology, and background. For scale, converting representative flux-density sensitivities to integrated line-flux sensitivities for a $5000~\mathrm{km~s^{-1}}$ Gaussian gives $\sim1.1\times10^{-2}$--$1.1\times10^{-1}$ mJy~$\mu$m near $6~\mu$m for a $\sim0.1$--$1$ mJy MIRI/MRS threshold, and $\sim4.4\times10^{-5}$--$4.4\times10^{-4}$ mJy~$\mu$m near $2.5~\mu$m for a favorable $\sim10^{-3}$--$10^{-2}$ mJy NIR threshold. The predicted MIR lines remain many orders of magnitude below these values, while even the brightest NIR high-ionization lines remain at least several hundred to several thousand times too faint.
    These results demonstrate that even high-ionization lines formed at small radii do not produce detectable variability on hour timescales. The lack of detectable rapid variability reflects the fact that the recombination and cooling timescales of the gas exceed the $\sim$hour duration of the flare, preventing the ionization state from responding impulsively to the injected radiation.}
    \label{fig:mir_nir_lc}
\end{figure*}

This indicates that the absence of detectable variability is not simply a consequence of the MIRI bandpass, but instead reflects a general property of the circumnuclear plasma. Even in the most favorable regime (for example, [\ion{Si}{9}] reaches $r \sim 10^{2}~R_g$), the gas responds primarily to the time-averaged ionizing field, effectively smoothing over rapid fluctuations. Figure~\ref{fig:mir_nir_lc} should therefore not be interpreted as showing an absence of fractional variability. Rather, it shows that the modeled fractional variability occurs at absolute flux densities far below the sensitivity of \textit{JWST}. The limiting factor is the combination of faint emissivity, long response times, overionization at high luminosities, and poor contrast against the continuum. Consequently, both MIR and NIR coronal lines are ineffective tracers of hour-timescale flare-driven variability from \SgrA.

\begin{table*}
\centering
\footnotesize
\setlength{\tabcolsep}{2pt}
\caption{Physical conditions at the radius of peak emissivity for each ion in the \texttt{CLOUDY} models.}
\begin{tabular}{lcccccccc}
\hline\hline
Line & $\lambda$ ($\mu$m) & IP (eV) & Peak Flux (mJy) & $R_{\rm peak}$ ($R_g$) & $T_{\rm peak}$ (K) & $\alpha_R$ (cm$^3$ s$^{-1}$) & $t_{\rm rec}$ (days) & $t_{\rm cool}$ (days) \\
\hline
{[Si VI]}   & 1.96 & 166.77 & $8.64\times10^{-3}$ & 18041 & 12992 & $7.20\times10^{-12}$ & $9.17\times10^{3}$ & $1.78\times10^{6}$ \\
{[Si VII]}  & 2.48 & 205.05 & $5.13\times10^{-2}$ & 7612  & 17277 & $1.88\times10^{-11}$ & $1.48\times10^{3}$ & $9.96\times10^{5}$ \\
{[Si IX]}   & 2.58 & 303.17 & $3.50\times10^{-2}$ & 330   & 52480 & $1.52\times10^{-10}$ & $7.73$ & $1.28\times10^{5}$ \\
{[Al V]}    & 2.91  & 120    & $1.07\times10^{-4}$ & 46247 & 9270  & $1.26\times10^{-12}$ & $1.34\times10^{5}$ & $3.25\times10^{6}$ \\
{[Mg VIII]} & 3.03  & 224.95 & 0.393      & 1291  & 29771 & $2.33\times10^{-11}$ & $2.03\times10^{2}$ & $2.91\times10^{5}$ \\
{[K VII]}   & 3.19 & 99.4   & $3.07\times10^{-4}$ & 3088  & 22391 & $2.28\times10^{-10}$ & $4.96\times10^{1}$ & $5.24\times10^{5}$ \\
{[Ca IV]}   & 3.21  & 50.91  & $1.31\times10^{-6}$ & 48084 & 9184  & $3.96\times10^{-12}$ & $4.44\times10^{4}$ & $3.35\times10^{6}$ \\
{[Al VI]}   & 3.66   & 153.83 & $5.56\times10^{-3}$ & 21950 & 12002 & $1.26\times10^{-12}$ & $6.38\times10^{4}$ & $2.00\times10^{6}$ \\
{[Al VIII]} & 3.69   & 241.44 & $9.39\times10^{-3}$ & 1291  & 29771 & $1.55\times10^{-11}$ & $3.04\times10^{2}$ & $2.91\times10^{5}$ \\
{[Si IX]}   & 3.94 & 303.17 & 1.51      & 392   & 48394 & $1.52\times10^{-10}$ & $7.92$ & $1.21\times10^{5}$ \\
{[Ca VII]}  & 4.09  & 108.78 & $8.25\times10^{-6}$ & 2100  & 25197 & $8.55\times10^{-12}$ & $8.99\times10^{2}$ & $4.01\times10^{5}$ \\
{[Ca V]}    & 4.16 & 67.27  & $1.44\times10^{-5}$ & 44459 & 9352  & $7.08\times10^{-12}$ & $2.30\times10^{4}$ & $3.15\times10^{6}$ \\
{[Mg IV]}   & 4.49  & 80.14  & $6.45\times10^{-3}$ & 48084 & 9184  & $3.50\times10^{-12}$ & $5.03\times10^{4}$ & $3.35\times10^{6}$ \\
{[Ar VI]}   & 4.53  & 75.02  & 0.293     & 9262  & 16303 & $4.91\times10^{-11}$ & $6.90\times10^{2}$ & $1.14\times10^{6}$ \\
{[Na VII]}  & 4.68  & 172.15 & 0.0968    & 2100  & 25197 & $1.20\times10^{-11}$ & $6.41\times10^{2}$ & $4.01\times10^{5}$ \\
\hline
{[Mg VII]}  & 5.50 & 186.76 & 0.425      & 2554  & 22152 & $7.33\times10^{-11}$ & $1.28\times10^{2}$ & $4.29\times10^{5}$ \\
{[K VI]}    & 5.57 & 82.66  & $1.58\times10^{-3}$ & 4975  & 18338 & $1.15\times10^{-10}$ & $1.58\times10^{2}$ & $6.91\times10^{5}$ \\
{[Mg V]}    & 5.61  & 109.27 & 0.198     & 29021 & 9944  & $7.70\times10^{-12}$ & $1.38\times10^{4}$ & $2.19\times10^{6}$ \\
{[Al VIII]} & 5.85 & 241.44 & $4.28\times10^{-3}$    & 1026  & 29786 & $1.55\times10^{-11}$ & $2.42\times10^{2}$ & $2.31\times10^{5}$ \\
{[Ca VII]}  & 6.15 & 108.78 & $2.69\times10^{-5}$ & 1669  & 25206 & $8.55\times10^{-12}$ & $7.15\times10^{2}$ & $3.19\times10^{5}$ \\
{[Si VII]}  & 6.51 & 205.05 & $1.01\times10^{-2}$    & 5820  & 17535 & $1.88\times10^{-11}$ & $1.13\times10^{3}$ & $7.73\times10^{5}$ \\
H I     & 7.46 & 13.6   & $6.75\times10^{-5}$ & 48084 & 9184  & $2.66\times10^{-13}$ & $6.62\times10^{5}$ & $3.35\times10^{6}$ \\
{[Ne VI]}   & 7.65 & 126.25 & 6.00      & 7365  & 16364 & $1.24\times10^{-11}$ & $2.18\times10^{3}$ & $9.13\times10^{5}$ \\
{[Fe VII]}  & 7.81 & 99.1   & $3.83\times10^{-3}$    & 15518 & 12675 & $2.22\times10^{-11}$ & $2.56\times10^{3}$ & $1.49\times10^{6}$ \\
{[Ar V]}    & 7.90 & 59.81  & $1.16\times10^{-3}$     & 18154 & 11882 & $3.91\times10^{-11}$ & $1.70\times10^{3}$ & $1.63\times10^{6}$ \\
{[K VI]}    & 8.82 & 82.66  & $3.26\times10^{-3}$ & 5174  & 18129 & $1.15\times10^{-10}$ & $1.64\times10^{2}$ & $7.11\times10^{5}$ \\
{[Na IV]}   & 9.03  & 71.62  & $1.63\times10^{-3}$ & 46464 & 8843  & $3.91\times10^{-12}$ & $4.35\times10^{4}$ & $3.11\times10^{6}$ \\
{[Mg VII]}  & 9.03  & 186.76 & 5.41      & 2554  & 22152 & $7.33\times10^{-11}$ & $1.28\times10^{2}$ & $4.29\times10^{5}$ \\
{[Al VI]}   & 9.11 & 153.83 & $2.46\times10^{-3}$ & 17456 & 12078 & $2.94\times10^{-12}$ & $2.17\times10^{4}$ & $1.60\times10^{6}$ \\
{[Fe VII]}  & 9.51 & 99.1   & 0.0656    & 15518 & 12675 & $2.22\times10^{-11}$ & $2.56\times10^{3}$ & $1.49\times10^{6}$ \\
{[Ca V]}    & 11.48& 67.27  & $5.92\times10^{-6}$ & 33951 & 9491  & $7.08\times10^{-12}$ & $1.76\times10^{4}$ & $2.44\times10^{6}$ \\
{[Ne II]}   & 12.81 & 21.56  & 0.0898    & 49851 & 9107  & $1.49\times10^{-13}$ & $1.22\times10^{6}$ & $3.44\times10^{6}$ \\
{[Ar V]}    & 13.10& 59.81  & 0.688     & 19636 & 11502 & $3.91\times10^{-11}$ & $1.84\times10^{3}$ & $1.71\times10^{6}$ \\
{[F V]}     & 13.43& 87.14  & $5.69\times10^{-6}$ & 9692  & 15011 & $2.36\times10^{-11}$ & $1.50\times10^{3}$ & $1.10\times10^{6}$ \\
{[Mg V]}    & 13.55& 109.27 & 0.0388    & 27905 & 10072 & $7.70\times10^{-12}$ & $1.33\times10^{4}$ & $2.13\times10^{6}$ \\
{[Na VI]}   & 14.39& 138.4  & 0.110     & 7082  & 16557 & $8.66\times10^{-11}$ & $2.99\times10^{2}$ & $8.88\times10^{5}$ \\
{[Ne V]}    & 14.32& 97.12  & 11.2      & 26831 & 10207 & $6.43\times10^{-12}$ & $1.53\times10^{4}$ & $2.07\times10^{6}$ \\
{[P III]}   & 17.89 & 19.77  & 0.754     & 48084 & 9184  & $9.07\times10^{-12}$ & $1.94\times10^{4}$ & $3.35\times10^{6}$ \\
{[Fe II]}   & 17.94 & 7.9    & $3.23\times10^{-4}$ & 65472 & 8629  & $7.20\times10^{-13}$ & $3.33\times10^{5}$ & $4.28\times10^{6}$ \\
{[S III]}   & 18.71 & 23.34  & $2.78\times10^{-8}$ & 48084 & 9184  & $3.77\times10^{-12}$ & $4.67\times10^{4}$ & $3.35\times10^{6}$ \\
{[Na IV]}   & 21.32& 71.62  & $3.18\times10^{-4}$ & 46464 & 8843  & $3.91\times10^{-12}$ & $4.35\times10^{4}$ & $3.11\times10^{6}$ \\
{[Ne V]}    & 24.31& 97.12  & 23.7      & 27905 & 10072 & $6.43\times10^{-12}$ & $1.59\times10^{4}$ & $2.13\times10^{6}$ \\
{[F IV]}    & 25.77& 62.71  & $2.40\times10^{-4}$ & 46464 & 8843  & $2.67\times10^{-11}$ & $6.37\times10^{3}$ & $3.11\times10^{6}$ \\
{[O IV]}    & 25.88& 54.94  & 18.1      & 46464 & 8843  & $1.41\times10^{-11}$ & $1.21\times10^{4}$ & $3.11\times10^{6}$ \\
\hline
\end{tabular}
\label{tab:nir_timescales}
\raggedright
\tablecomments{For each transition, we report the wavelength, ionization potential (IP), peak flux (mJy), radius of peak emission ($R_{\rm peak}$), gas temperature ($T_{\rm peak}$), recombination coefficient ($\alpha_R$), and the corresponding recombination and cooling timescales. The transitions are ordered by increasing wavelength. The recombination coefficients are taken from \citet{Nahar1995}, \citet{ADAS}, or \texttt{ChiantiPy} \citep{ChiantiPy}. The recombination times are calculated as $t_{\rm rec} = (\alpha_R n_e)^{-1}$ using the local electron density at $R_{\rm peak}$, while the cooling times are estimated using a representative cooling coefficient of $\Lambda = 10^{-22}$ erg cm$^{3}$ s$^{-1}$. Across all ions, both the recombination and cooling timescales significantly exceed the typical duration of \SgrA\ flares ($\sim$1 hour), even for high-ionization species that originate at smaller radii and higher densities (e.g., Si~IX). The upper section lists near-infrared lines; the lower section lists mid-infrared lines.}
\end{table*}

% Figure~\ref{fig:mir_nir_lc} shows the predicted line lightcurves for representative high-ionization MIR and NIR transitions at the highest flare luminosity tested at $\log L_{\mathrm{flare}} = 39$. We find that the vast majority of emission lines exhibit extremely low observed fluxes, typically spanning $\sim 10^{-17}$--$10^{-7} \ \mathrm{mJy}$ even before extinction is applied. These values lie far below the continuum sensitivity limits of the \textit{JWST}/MIRI observations and are therefore intrinsically undetectable. In both wavelength regimes, the emission remains extremely faint and exhibits only weak temporal evolution, with fluxes either declining or remaining approximately constant after the flare injection. This behavior reflects the fact that even for lines formed at smaller radii, the recombination and cooling timescales exceed the flare duration, suppressing any rapid variability.

\section{Discussion} \label{sec:discussion}

\begin{figure*}
    \centering
    \includegraphics[width=0.98\textwidth]{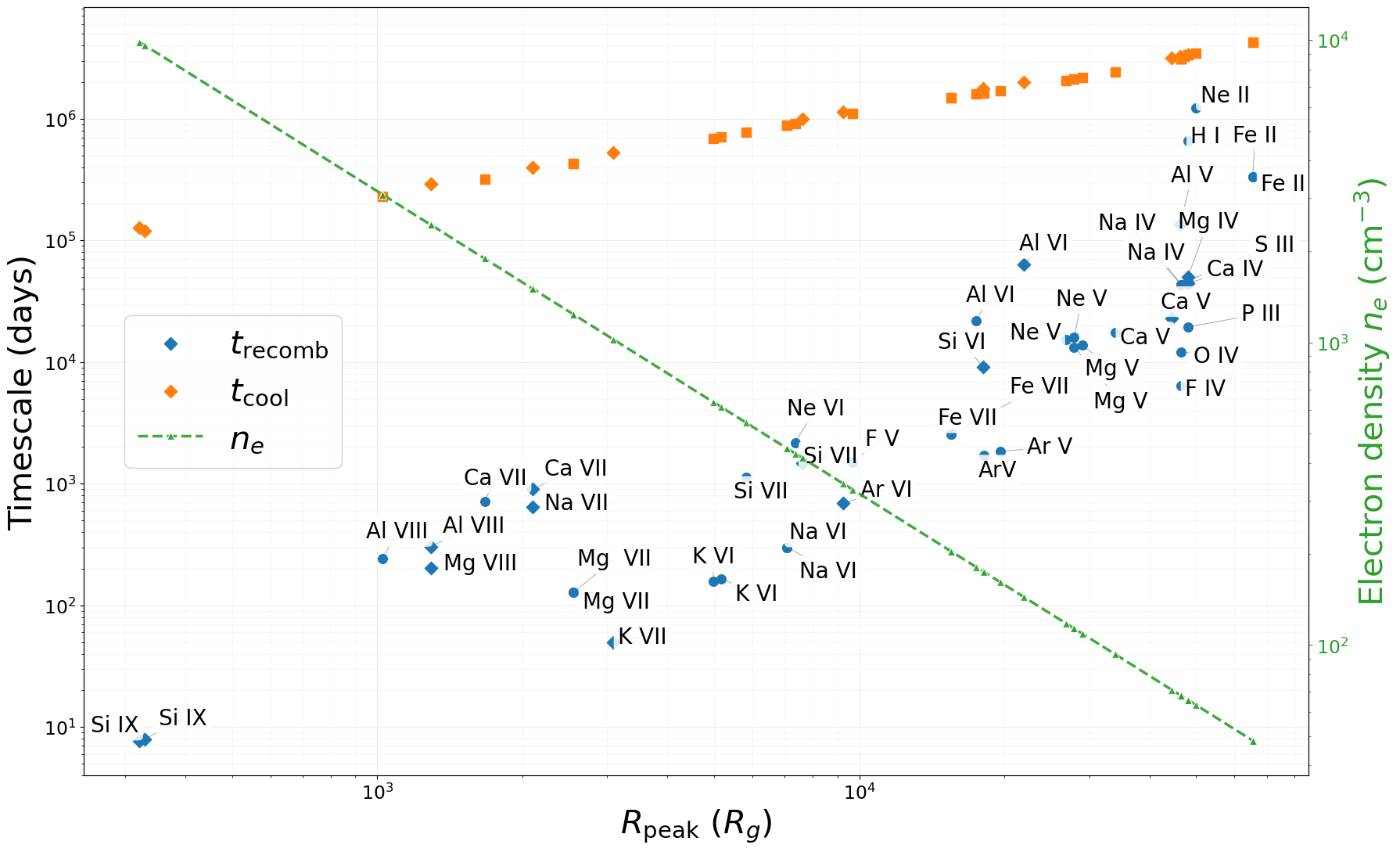}
    \caption{Recombination time (blue points), cooling time (orange squares; computed assuming a representative cooling coefficient of $10^{-22}$ erg cm$^{3}$ s$^{-1}$), and density (green diamonds) as a function of the radius at which each ion reaches peak emissivity. Ion labels are shown above the recombination-time points. Even the shortest recombination timescale is of order $\sim8$ days, far exceeding the characteristic variability timescale of $\sim1$ hour. }
    \label{fig:tcool}
\end{figure*}

The absence of detectable continuum-correlated variability in the \textit{JWST}/MIRI emission-line light curves reflects how circumnuclear gas around \SgrA\ reprocesses a rapidly varying radiation field. The multi-method uncertainty analysis shows that the observed fluctuations are consistent with measurement scatter, and the resulting $L_{\rm var,max}$ values place upper limits on any hidden variable component. Physically, the line-emitting gas is spatially extended and thermodynamically slow to respond, so hour-timescale flares are smoothed into a quasi-steady emission component.

The geometry of the emitting region is a key limitation. The observed line flux integrates gas over a wide range of radii, combining zones with different light-travel times. For a flare of duration $t_{\rm flare}$, only gas within $r_{\rm lc}=ct_{\rm flare}\approx180 R_g$ can respond coherently. \texttt{CLOUDY} emissivity profiles show that the fraction of emission from this region is negligible, $f(r_{\rm lc})\lesssim10^{-7}$. Most emission arises from radii with light-crossing times of $\sim$0.1--10 days, diluting any rapid response.

This disparity between the monitoring duration and the light-crossing time of the full aperture is important. Gas at larger radii would respond on day-to-week timescales, beyond the observations, and its emission would be further diluted by integration over many shells. Combined with long recombination and cooling times, any delayed response is broadened into the quasi-steady component rather than producing a coherent flare signal. The non-detection therefore reflects the absence of a prompt response, not all delayed variability within the aperture.

Thermodynamic constraints reinforce this picture. Figure~\ref{fig:tcool} shows recombination and cooling times at peak emissivity radii. Even for compact tracers, these timescales far exceed the $\sim$hour flare duration. For example, [\ion{Al}{8}] peaks at $\sim10^{3} R_g$ with a recombination time of $\sim242$ days, while [\ion{Si}{9}] peaks at $\sim330 R_g$ with $\sim8$ days. In both cases, the gas cannot adjust on flare timescales and instead reflects the time-averaged radiation field.

We tested the sensitivity of the predictions to density by increasing the normalization from
$n=10^{6.5}$ to $10^{7.5}~\mathrm{cm^{-3}}$ at $10R_g$. Although this test does not change our conclusion that the detected MIRI lines are ineffective probes of hour-timescale flares, it produces a potentially useful independent density diagnostic. The principal new MIR feature is [\ion{Mg}{7}] $5.502~\mu$m, which reaches peak emissivity at
$r_{\rm Mg\ VII} \approx 2.6\times10^{3}R_g$ but is not detected in the MIRI/MRS spectra. In our adopted $n\propto r^{-1}$ profile, the high-density model corresponds to a local density of order $\sim10^{5}~\mathrm{cm^{-3}}$ at the radius of peak emissivity. The absence of the line therefore disfavors densities substantially above this scale, conditional on the assumed temperature structure, abundances, and flare spectrum. Because the calculation does not include a complete geometric delay distribution or transfer function, we regard this as an order-of-magnitude density constraint rather than a precise upper limit. This illustrates that, although MIR lines do not constrain the instantaneous UV luminosity of \SgrA, their presence or absence may provide an independent probe of the density structure of the accretion flow at specific radii.

The mismatch between observed MIRI lines and model predictions is informative. The detected [\ion{Fe}{2}], [\ion{Ne}{2}], and [\ion{S}{3}] lines likely arise from cooler, denser, and more extended circumnuclear gas, such as Sgr A West, rather than the hot accretion flow. These regions can produce bright MIR emission under steady radiation fields without requiring flare excitation. The \texttt{CLOUDY} models instead isolate the flare response of hot gas, which is intrinsically weak and temporally smeared. The continuum-dominated spectra reinforce this conclusion. Increasing $L_{\mathrm{flare}}$ mainly raises the continuum rather than producing strong lines, so detectability is limited by line-to-continuum contrast. Even without smearing, the intrinsic line contrast is low, making flare-driven components difficult to isolate.

Kinematic broadening further reduces detectability. At $r\sim10^{3} R_g$, velocities of $\sim10^{3}~\mathrm{km~s^{-1}}$ produce $\mathrm{FWHM}\sim10^{3}~\mathrm{km~s^{-1}}$, spreading emission over $\Delta\lambda/\lambda\sim10^{-3}$ and lowering peak contrast. Combined with low emissivity and continuum dilution, this suppresses any variable signal, consistent with \texttt{CLOUDY} simulations showing no prompt response.

This places the Galactic Center in a reverberation-mapping context distinct from classical AGN. Broad-line regions respond rapidly to continuum changes, while narrow-line regions reflect time-averaged luminosities due to longer timescales \citep{Blandford1982, Peterson2004, Bentz2015, Sun2018, Dempsey2018}. Extended emission-line regions similarly trace long-term activity \citep{Schirmer2016}. Although the gas here lies at smaller radii ($r\sim10^{2}$--$10^{5} R_g$), its behavior is governed by the same interplay of density, light-crossing time, and recombination physics. Because response times exceed flare durations, the gas acts as a low-pass filter, preserving the time-averaged radiation field and suppressing individual flares. The key factor is whether the gas can respond on the variability timescale, not its absolute radius.

Extending to higher-ionization MIR and NIR lines does not change the conclusion. These lines originate closer in but remain faint and slow to respond. Even in favorable cases, predicted fluxes are $\lesssim10^{-7}~\mathrm{mJy}$, far below detectability, and do not show observable rapid variability. The limitations arise from faint emissivity, long response times, overionization, kinematic broadening, and poor contrast.

Overall, infrared emission lines are poor diagnostics of the instantaneous UV luminosity of \SgrA\ flares. The lack of variability does not imply weak UV emission, but rather that the gas cannot translate rapid changes into detectable line variability. The observed MIRI lines trace quasi-steady circumnuclear material, while flare-excited emission from hot gas is too weak and smeared to isolate. Stronger constraints will require probes of smaller radii, higher densities, or lower continuum backgrounds, where response times and contrast are more favorable.

\section{Conclusions} \label{sec:conclusions}

We investigate the response of circumnuclear gas to \SgrA\ flares using \textit{JWST}/MIRI spectroscopy and \texttt{CLOUDY} modeling. Our main conclusions are:

\begin{enumerate}
\item We find no variability in any of the observed mid-infrared emission lines (e.g., [\ion{Fe}{2}] 5.34~$\mu$m, [\ion{Ne}{2}] 12.813~$\mu$m, [\ion{Fe}{2}] 17.936~$\mu$m, [\ion{S}{3}] 18.713~$\mu$m) that correlates with the continuum. The small fluctuations we do see are consistent with measurement noise. Across the observed lines and epochs, the most conservative limit on the measured variability is $\langle \nu L_\nu \rangle = (7.8 \pm 0.8)\times 10^{33}$~\ergs. 
\item The lack of line variability is expected given the structure of the gas. The emission comes from a wide range of distances, with light-travel times of about 0.1–10 days. Only a tiny fraction of the emission comes from regions that could respond quickly (on hour timescales), so any rapid signal is smeared out when summed over all radii.
\item The gas itself also responds slowly. The models show that recombination and cooling times are much longer than typical flare durations, even for highly ionized species. As a result, the gas mainly reflects the average radiation field rather than individual flares.
\item Extending the analysis to prospective observations, we find that high-ionization MIR and NIR emission lines are intrinsically faint and remain far below current \textit{JWST} sensitivity limits. While such lines may in principle be detectable with future instruments offering orders-of-magnitude improvements in sensitivity, observations of infrared lines currently cannot constrain the instantaneous UV emission of \SgrA.
\end{enumerate}

These results demonstrate that the circumnuclear medium around \SgrA\ is insensitive to short-timescale variability and instead traces the steady-state radiation field.

\begin{acknowledgments}
All of the data presented in this article were obtained from the Mikulski Archive for Space Telescopes (MAST) at the Space Telescope Science Institute. The specific observations analyzed can be accessed via \dataset[DOI: 10.17909/580m-xg23]{https://doi.org/10.17909/580m-xg23}.

M.B. thanks Steven P. Willner for very beneficial discussions. M.B. also thanks Hannah Dykaar and Sophia Sanchez-Maes for helpful conversations and acknowledges support from the Natural Sciences and Engineering Research Council of Canada's Banting Postdoctoral Fellowship Program (CIHR AWARD BPF 200617-267964). D.H., Z.S., N.M.F., M.B. and S.D.vF. acknowledge support from the Canadian Space Agency under awards 23JWGO2A01 and 25JWGO4A01. D.H., Z.S., N.M.F. and M.B. also acknowledge funding from the Natural Sciences and Engineering Research Council of Canada (NSERC) Arthur B. McDonald Fellowship and Discovery Grant programs, the Canada Research Chairs (CRC) program, the Fondes de Recherche Nature et Technologies (FRQNT), the Centre de recherche en astrophysique du Québec (un regroupement stratégique du FRQNT), and the Trottier Space Institute at McGill. 

N.M.F. acknowledges funding from the FRQNT Doctoral Research Scholarship and NSERC Canada Graduate Research Scholarship. L.L. acknowledges the support of DGAPA-PAPIIT grant IN108324, CONAHCyT grant CF-263356, and SECIHTI grant CBF-2025-I-109. S.D.vF. gratefully acknowledges the support of the Alexander von Humboldt Foundation through a Feodor Lynen Fellowship and thanks CITA for their hospitality and collaboration
S.D.vF. acknowledge the support of the Natural Sciences and Engineering Research Council of Canada (NSERC), [funding reference number 568580] Cette recherche a \'et\'e financ\'ee par le Conseil de recherches en sciences naturelles et en g\'enie du Canada (CRSNG), [num\'ero de r\'ef\'erence 568580]. J.M.M. is supported by an NSF Astronomy and Astrophysics Postdoctoral Fellowship under award AST-2401752.

This manuscript was prepared with the assistance of AI-based large language models (Claude, GPT-5 family) for language editing and Python code generation. The authors take full responsibility for the scientific content, data analysis, and interpretation presented in this work. All code was reviewed, validated, and executed by the authors.

This work is based [in part] on observations made with the NASA/ESA/CSA James Webb Space Telescope. 
The data were obtained from the Mikulski Archive for Space Telescopes at the Space Telescope Science Institute, which is operated by the Association of Universities for Research in Astronomy, Inc., under NASA contract NAS 5-03127 for JWST. These observations are associated with program \#4572 and \#7532.
The observations are available at the Mikulski Archive for Space Telescopes (\url{https://mast.stsci.edu/}). Support for program \#4572 was provided by NASA through a grant from the Space Telescope Science Institute, which is operated by the Association of Universities for Research in Astronomy, Inc., under NASA contract NAS 5-03127.

\end{acknowledgments}

%\begin{contribution}
%%This section gives authors the space to recognize author contributions. The text inside this environment is NOT counted towards the total word quanta. At a minimum, manuscripts are expected to include this text:

%% But authors are expected to provide more specific details, e.g. 
%%
%%SC was responsible for writing and submitting the manuscript.
%%WWM came up with the initial research concept and edited the manuscript.
%%OTS obtained the funding and edited the manuscript.
%%EBF provided the formal analysis and validation. He also edited the manuscript.
%%GEH Supervised the undergraduates, wrote the software and administers the project github and Zenodo repositories.
%%
%% Authors can use the Contributor Role Taxonomy (CRediT) at
%% for ideas on how write a good statement tailored to their needs.

%\end{contribution}

\facilities{JWST (MIRI)}

\software{Astropy \citep{astropy:2013, astropy:2018, astropy:2022}, Cloudy \citep{Gunasekera2025}, Jdaviz \citep{jdaviz}, {MRS Light Curve Calibration Pipeline}
\citep{vonFellenberg2025,Michail2026},
\href{https://doi.org/10.5281/zenodo.19558572}
{\texttt{JWST} Calibration Pipeline v2.0.0}
\citep{bushouse_2026_19558572}
          }


\appendix

\section{Search for Continuum-correlated Line Variability} \label{app:crosscorr}

We perform a complementary search for delayed continuum-correlated line variability. For each emission line, we compare the continuum-subtracted emission-line light curve, $L(t)$, to the contemporaneous mid-infrared continuum light curve, $C(t)$, extracted from the same MIRI channel.  We search for a delayed relationship between the line and continuum light curves over the lag range accessible to each observation (over positive-lag windows extending to 1 hr for the 2024 observations and 8 hr for the 2025 observations), using the interpolated cross-correlation function \citep[ICCF;][]{Gaskell1987,Peterson1993,Peterson2004}. At each trial lag $\tau$, one light curve is shifted relative to the other and linearly interpolated onto the sampling times of the unshifted light curve. We then compute the Pearson correlation coefficient between the overlapping points. Repeating this calculation over a grid of trial
lags produces the ICCF, $r(\tau)$. A positive value of $r(\tau)$ indicates that similar continuum and line-flux variations occur with a delay $\tau$. Because a reprocessing response should causally follow the driving continuum variability, we restrict the search to positive lags ($L(t)$ is correlated with $C(t-\tau)$), using $0<\tau<60$ min for the shorter 2024 observations and $0<\tau<480$ min for the longer 2025 observations.

The significance of the maximum observed correlation coefficient, $r_{\rm max}$, is evaluated using null light curves that preserve the temporal structure of the data but remove any physical continuum--line correlation. Specifically, we generate phase-randomized surrogate line light curves and repeat the ICCF analysis on each realization. The 95th percentile of the resulting null distribution defines the critical correlation coefficient, $r_{\rm crit,95}$; a peak in the ICCF is considered formally significant only if $r_{\rm max}>r_{\rm crit,95}$.

To translate the non-detections into limits on a reverberating response, we perform injection--recovery tests using the observed continuum light curve as the driving signal. For each realization, we construct a synthetic delayed line light curve,
\begin{equation}
L_{\rm inj}(t) =
\eta L_{\rm ref}
\frac{C(t-\tau_{\rm inj})-\langle C\rangle}
{\langle C\rangle}
+\epsilon(t),
\end{equation}
\label{eq:inj}
where $\tau_{\rm inj}$ is the injected lag and $\epsilon(t)$ is Gaussian noise drawn independently at each integration using the corresponding measured line-flux uncertainty. The reference scale $L_{\rm ref}$ sets the characteristic line-flux amplitude used in the injection test. It is taken to be the mean absolute value of the measured continuum-subtracted line flux for that transition and epoch. The dimensionless parameter $\eta$ sets the amplitude of the injected response. Following Equation \ref{eq:inj}, $\eta=1$ produces a delayed line-flux change equal to $L_{\rm ref}$.

We test 25 linearly spaced response amplitudes over $0.05\leq\eta\leq5.0$. For each amplitude, we generate 300 realizations with $\tau_{\rm inj}$ drawn from the same positive-lag grid searched in the observed data. We calculate the ICCF of the observed continuum and each synthetic line light curve and classify a response as recovered when its maximum correlation coefficient satisfies $r_{\rm max}\geq r_{\rm crit,95}$.

Across the full set of lines and epochs, we find no physically compelling continuum-correlated line response. Two cases formally exceed the 95\% null threshold: [\ion{Fe}{2}] $5.34~\mu$m in jw04572003001, with $r_{\rm max}=0.87$ at $\tau=+60$ min compared to $r_{\rm crit,95}=0.85$ ($p=0.017$), and [\ion{Ne}{2}] $12.81~\mu$m in jw07532004001, with $r_{\rm max}=0.99$ at $\tau=+415$ min compared to $r_{\rm crit,95}=0.98$ ($p=0.015$). We do not interpret either case as a robust reverberation detection. Both maxima occur at or near the upper boundary of the searched lag interval: the [\ion{Fe}{2}] peak is pinned to the $+60$ min edge of the 2024 search window, while the [\ion{Ne}{2}] peak occurs at nearly the full 2025 visit baseline. At such large lags, shifting one light curve relative to the other leaves progressively fewer overlapping integrations, making the inferred correlation increasingly sensitive to the observing window and temporal sampling. Previous simulation-based studies have shown that ICCF lag recovery can become unreliable under these conditions, particularly when the lag approaches a poorly sampled portion of the available baseline \citep{Li2019,Malik2022,McDougall2025}. Because neither maximum is localized within the interior of the searched lag range, we treat these cases as boundary-lag candidates rather than robust reverberation detections.

We therefore treat these formal peaks as non-detections and report the result as an upper limit on continuum-correlated line variability. The most conservative constraint comes from our noisiest transition, [\ion{S}{3}] $18.713~\mu$m. For this line, no continuum-correlated variability is detected, and any reverberating response stronger than $\approx459\%$ of the reference line flux over the positive-lag range $0<\tau<480$ min is excluded at 95\% confidence. This supports the interpretation that the line-emitting gas responds primarily to the time-averaged radiation field rather than to individual hour-timescale flares.

\bibliography{sample701}{}
\bibliographystyle{aasjournalv7}



\end{document}